\documentclass[10pt]{article}            
\usepackage[colorlinks=true,breaklinks=true,bookmarks=true,urlcolor=blue,
     citecolor=blue,linkcolor=blue,bookmarksopen=false,draft=false]{hyperref}
        
\usepackage{authblk}
\usepackage{amsmath}
\usepackage{amssymb}
\usepackage{amsthm}
\usepackage{import}
\usepackage{esvect}
\usepackage{tikz}
\usetikzlibrary{matrix}
\usepackage{graphicx}
\usepackage{pgfplots}
\pgfplotsset{compat=newest}
\usepackage{caption}
\usepackage{subcaption}
\usepackage{algorithm}
\usepackage{algpseudocode}
\usepackage{enumitem}
\usepackage{ dsfont }
\usepackage{biblatex}
\setlist{nosep}
\newtheorem{definition}{Definition}
\newtheorem{axiom}{Axiom}
\newtheorem{theorem}{Theorem}
\newtheorem{theorem*}{Theorem}
\newtheorem{lemma}{Lemma}
\newtheorem{corollary*}{Corollary}
\newtheorem{remark}{Remark}
\newtheorem{axiom*}{Axiom}
\newtheorem{proposition}{Proposition}
\newtheorem{proposition*}{Proposition}

\begin{document}

\title{Level-Strategypoof Belief Aggregation and Application to Majority Judgment under Uncertainty}

\author[1]{Rida Laraki}
\author[2]{Estelle Varloot}
\affil[1]{CNRS (LAMSADE, Université Paris Dauphine-PSL),University of Liverpool}
\affil[2]{University of Liverpool}
\affil[]{rida.laraki@dauphine.fr, estelle.varloot@liverpool.ac.uk}
\date{}

\maketitle

\begin{abstract}
In the problem of aggregating experts' probabilistic predictions or opinions over an ordered set of outcomes, we introduce the axiom of level-strategyproofness (Level-SP) and argue that it is natural in several real-life applications and robust as a notion. It implies truthfulness in a rich domain of single-peaked preferences over the space of cumulative distributions. This contrasts with the existing literature, where we usually assume single-peaked preferences over the space of probability distributions. Our main results are (1) explicit characterizations of all Level-SP methods with and without the addition of other axioms (certainty preservation, plausibility preservation, proportionality, weighted voters); (2) comparisons and axiomatic characterizations of two new and practical Level-SP methods: the proportional-cumulative and the middlemost-cumulative; (3) a use of the proportional-cumulative to construct a new voting method (MJU) that extends majority judgment (MJ) method for electing and ranking alternatives. In MJ, a voter attributes a grade on a scale of merits such as $\Lambda=\{$ Great; Good; Average; Poor; Terrible$\}$ to each candidate/alternative. In MJU, voters can express their uncertainties/doubts about the merits (e.g. a voter can submit, for each candidate, a probability distribution over $\Lambda$). We show that MJU inherits most of the salient properties of MJ (e.g. it avoids Arrow and Condorcet paradoxes and it resists to some natural strategic manipulations).
\end{abstract}

\section{Introduction.}\label{Introduction}

The main focus of this paper is the aggregation of experts' probabilistic opinions in an incentive-compatible way, without money transfers. It axiomatically characterizes new voting methods where reporting their preferred output is the optimal strategy for every expert, for a large set of utility functions.

In many real-life situations, even the most prominent experts are uncertain ---their opinions or predictions are probabilistic--- and may disagree in their judgments, even if they share a common interest with the regulator \cite{TBK0}. Thus, a method is needed to pool their opinions.

For example, with no evidence-based information on the Covid-19 disease, the European Academy of Neurology (EAN) developed an ad-hoc three-round voting method\footnote{``In round 1, statements were provided by EAN scientific panels (SPs). In round 2, these statements were circulated to SP members not involved in writing them, asking for agreement/disagreement. Items with agreement $>70\%$ were retained for round 3, in which SP co-chairs rated importance on a five-point Likert scale.Results were graded by importance and reported as consensus statements.''}  to reach a consensus \cite{Covid}.
When a potentially dangerous volcano becomes restless, civil authorities turn to scientists for help in anticipating risks. To do so, volcanologists developed elaborate mathematical models to elicit and aggregate experts' probabilistic opinions \cite{Volcano1,Volcano2}.
The Technical University of Delft has developed a software \hyperlink{https://www.tudelft.nl/ewi/over-de-faculteit/afdelingen/applied-mathematics/applied-probability/research/research-themes/risk/software/excalibur/excalibur}{EXCALIBUR}  and its successor ANDURYL \cite{LM2018}. 
This software is used in forecasting the weather, in calculating the risks to manned spaceflight due to collision with space debris, or in estimating the future of the polar bear population (see \cite{ATA,OOHLW} and chapter 15 in \cite{Cooke}).\footnote{We refer the reader to Roger Cooke web page: \url{http://rogermcooke.net} which contains useful references, real data, a wide range of applications and links to software.}

Aggregating experts' probabilistic opinions is a well-studied mathematical/social choice problem \cite{Armstrong1,French,MartiniS,Cooke}, sometimes referred to as belief aggregation or opinion pooling. The formal model is this: each expert $i\in N$ is asked to provide their regulator with their prior probability distribution $p_i\in \mathcal{P} =\Delta(\Lambda)$ over a set of outcomes $\Lambda$. The objective of the regulator is to design a PAF $=$ Probability Aggregation Function $\psi:\mathcal{P}^N \rightarrow \mathcal{P}$ satisfying some desirable properties. 

Most of the literature described so far assumes honest reporting by the experts of their desired outcome. In practice, judges may have strategic incentives. 

For example, the FDA  uses advisory committees and follows their recommendations 70\% of the time. There have been several controversies due to conflicts of interests within advisory committees \cite{elster_2015}. Hence, we wish for the PAF to be incentive compatible (IC), e.g. reporting its preferred output is an optimal strategy for every expert.

When money transfers are possible (e.g. experts can be paid after some realizations of the random variable), the problem has been well studied and several incentive compatible ``scoring rules'' have been designed \cite{Good52,McCarthy56,Winkler69,DGF83}. Our paper deals with situations where money transfers are impossible because the uncertain event studied is far in the future and/or the consequences of a  bad decision are potentially catastrophic (volcanic eruption, irreversible global warming, etc). 

To the best of our knowledge, incentive-compatible belief aggregation without money transfers has only been studied as a special case of single-peaked domain restrictions (and only when the set of outcomes $\Lambda$ is finite). First of all, for strategy\-proofness to be stated formally, one needs to make assumptions about the individual's preferences over the set $\mathcal{P}$ of probability distributions over  $\Lambda$. If the preferences are unrestricted, Gibbard \cite{G1973} and Satterthwaite's \cite{S1973} theorems apply and only a dictatorial method is strategy\-proof. Unfortunately, the Gibbard-Satterthwaite negative conclusion still holds even if one restricts the domain, assuming it is ``rich'' enough (e.g. the class of all convex preferences over $\mathcal{P}$ \cite{Z1991} or the class of generalized single-peaked preferences, including all additively separable convex preferences over $\mathcal{P}$ \cite{NP2007}). Fortunately, under a more severe restriction, the possibility of an anonymous aggregation has been proved recently \cite{GKSA2016,EC2019} in the domain of the $L_1$-metric single-peaked preferences on $\mathcal{P}$.\footnote{Single-peaked preferences on $\Delta(\Lambda)$ with $\Lambda$ finite under the $L_1$-metric is a very small domain because every peak corresponds exactly to one preference ordering.} As such, Goel et al. \cite{GKSA2016} proved the existence of a Pareto optimal strategy\-proof PAF and Freeman et al. \cite{EC2019} identified a large family of strategy\-proof PAF in the spirit of Moulin's characterization \cite{M1980} where phantom functions replace phantom voters. Nevertheless, as in \cite{GKSA2016,EC2019} the main motivating problem is not the probability but the budget aggregation problem,
their methods are neutral for $\Lambda$, that is, they are invariant to permutations of the elements of $\Lambda$. In many applications, neutrality is not a desirable property.

    Let us explain this last point with an example. Let us suppose that the outcome space $\Lambda$ is a finitely ordered set $\{a_m \succ\dots\succ a_2 \succ a_{1}\}$ (such as the Richter or the volcanic scale). To define the utility functions, we need to determine the cost for expert $i$ whenever the aggregate probability $p$ is different from its forecast $p_i$. The most commonly used method is distance measuring using some $L_q$-distance on $\mathcal{P}\subset \mathcal{R}^m$. For example \cite{GKSA2016,EC2019} uses the $L_1$-distance on $\mathcal{P}$. Given our context, this is not a practical measure. For example, imagine that $p_i$ (the peak of $i$) is the Dirac mass $\delta_{a_1}$ at the smallest alternative $a_1$ and that $p$ (the output) is the Dirac mass $\delta_{a_m}$ at the highest alternative $a_m\in \Lambda$. We would rather use a measure that states that from the expert $i$'s perspective, $\delta_{a_m}$ is the worst possible probability distribution. In particular, the Dirac mass at $a_2$ should be closer to $a_1$ than $a_m$. However, no $L_q$-distance on $\mathcal{P}=\Delta(\Lambda)\subset \mathcal{R}^m$ will distinguish $\|\delta_{a_1}-\delta_{a_m}\|_{L_q}$ from $\|\delta_{a_1}-\delta_{a_2}\|_{L_q}$ as both are equal (those metrics are neutral). 

To capture the notion that $\operatorname{dist}(\delta_{a_1},\delta_{a_m})$ should be bigger than  $\operatorname{dist}(\delta_{a_1},\delta_{a_2})$, it is natural to measure the distances in the space of cumulative distribution functions (CDFs) $\mathcal{C}= \Sigma(\Lambda)$. Characterizing strategy\-proof PAF when the preferences are single-peaked in the CDF space has, to our knowledge, never been done before. This is what our article does by fully characterizing all strategy\-proof methods under various combinations of axioms. 

\textbf{Our main contributions are:}

\begin{itemize}
    \item We define the new concept of level-strategy\-proofness (Level-SP). We prove it implies incentive compatibility (IC) for a rich class of single-peaked preferences in the space of CDFs.\footnote{Being IC for a rich class of preferences is a desirable robustness property in social choice theory. Unfortunately, it is very rarely satisfied.}
    \item We provide several characterizations of Level-SP methods with or without combinations with other axioms, and we explore their boundaries by establishing some impossibility results.
\item We characterize and compare two new methods: the middlemost and proportional cumulatives.
\item We use the proportional-cumulative to construct a new voting method (MJU) to rank alternatives where voters can express their uncertainties/doubts about the qualities/merits of each alternative.  MJU, in addition to giving a new right to the voters compared to majority judgment  \cite{BL2011} (MJ), it inherits many of MJ's salient properties (avoids Arrow and Condorcet paradoxes and resists to some strategic manipulations).
\end{itemize}

The intuition behind Level-SP is simple. Let us suppose the regulator's decision depends on the likelihood of crossing a certain threshold, for example, the probability of having a major natural hazard. Then, experts' incentives will also depend on the odds of crossing that threshold. If the aggregation rule is Level-SP, then, whatever the threshold considered, and even if it is not known of the regulator or the experts in advance, no expert can, by misreporting, influence the aggregated outcome for the threshold to obtain something closer to their truly desired value.

In addition to Level-SP, a desirable axiom is \textit{certainty preservation}. It says that if all experts agree that some event must happen (resp. cannot happen), then the aggregation preserves that property.\footnote{Formally, certainty preservation is defied as as follows: for every Borel measurable event $A$, if $p_i(A)=0$ for every $i\in N$ then $p(A)=0$ or equivalently if $p_i(A)=1$ for every $i\in N$ then $p(A)=1$.} This axiom is classical in the literature and is also called the \textit{zero preservation property} in \cite{MartiniS} or \textit{consensus preservation} in \cite{DL2017}. 
In addition, the regulator may want to satisfy the \textit{plausibility preservation} axiom: whenever all experts agree that some interval has positive probability,\footnote{If one asks this property for Borel measurable events, we reach an impossibility.} so does society. We prove that the order-cumulatives (e.g. the middlemost-cumulative) are the unique anonymous Level-SP PAFs that are certainty and plausibility preserving. Their main drawback is their lack of diversity: they are such that whenever the experts' inputs dominate one another, the output is one of the inputs (and not a combination of their opinions). 

Our second main method (we call the weighted-proportional-cumulative) solves this drawback. Supposing that the experts are given weights, whenever the inputs dominate one another, the weighted-proportional-cumulative agrees with each expert in proportion to their weight. It is characterized as the unique Level-SP method satisfying a ``proportionality axiom'', namely, if all inputs are the Dirac distributions $\{\delta_{a_i}\}_{i=1,\dots,n}$, then the output is their weighted average $\sum_{i=1}^n w_i \delta_{a_i}$ (where $w_i$ is the given weight of expert $i$). 

The paper is organized as follows. Section \ref{Model Context} introduces the model, the new notion of Level-SP, and its implications together with a quick summary of the fundamental results of Moulin \cite{M1980} in the one-dimensional framework.  Section \ref{characterization} characterizes all Level-SP methods in the general and anonymous cases.  Section \ref{section certainty} (resp. \ref{section plausibility}) isolates certainty (resp. plausibility) preserving Level-SP methods.
Section \ref{section combining} combines certainty preservation and plausibility preservation axioms and characterizes the middlemost-cumulative. Section \ref{section diversity} is dedicated to diversity, and it characterizes the proportional cumulative as the unique Level-SP method satisfying proportionality. Section \ref{section comparing} compares our new two cumulative methods. Section \ref{section weighted experts} models axiomatically what it means for experts to have weights and characterises all the weight consistent Level-SP methods. Section \ref{section MJ and MJU} uses the proportional cumulative to construct an extension of majority judgment \cite{BL2011} to electing and ranking problems where voters can express their uncertainties in their ballots (a voter can express in the ballot that candidate A would be a Great president with probability 60\% and Terrible with probability 40\%, and that candidate B would be an Average president for sure).   
%Section \ref{conclusion} concludes. 

\section{Models and concepts.}\label{Model Context}

We first recall the classical characterizations of strategy\-proof aggregation rules when voters have one-dimensional single-peaked preferences (Moulin \cite{M1980}). Then we describe our probability aggregation model. We introduce the notion of level-strategy\-proofness (Level-SP). We will then prove that it implies classical strategy\-proofness  for a rich family of single-peaked utility functions over the CDF space.

From here onwards, $N=\{1,\dots,n\}$ denotes the set of voters or experts, and $n$ their number. For any set $Z$, $\textbf{z} =(z_1,\dots,z_n)$ denotes an element of $Z^N$ (interpreted as a voting profile where each voter $i$ submits the input $z_i$). For a voting profile $\textbf{z} \in Z^N$, we let $\textbf{z}_{-i}(z'_i)$ denote the profile which differs from $\textbf{z}$ only in dimension $i$ which is replaced by $z'_i$. This is a standard notation in social choice theory and it corresponds to the notation $(z'_i,z_{-i})$ in game theory.

\subsection{Definition and characterization of one-dimensional strategyproofness.}

We recall Moulin's results \cite{M1980} and characterizations. They are essential to understanding our results.

\begin{definition}[Single-peaked preference]
A preference order $T$ (represented with $\preccurlyeq$) over the set of alternatives $[0,1]$ is \textbf{single-peaked} if there exists $x \in [0,1]$ such that  $\forall y,z \in [0,1]$, $z \leq y \leq x \Rightarrow z \preccurlyeq y$ and $x \leq y \leq z \Rightarrow z \preccurlyeq y$. The value $x$ is known as the \textbf{peak}.
\end{definition}

  In other words, an ordinal preference on the line $[0,1]$ is single-peaked iff the cardinal utility function representing it is weakly increasing until the peak, then it is weakly decreasing.

\begin{definition}[One-SP]
A voting rule $g : [0,1]^N \rightarrow [0,1]$ is a one-dimensional strategy\-proof
(\textbf{One-SP}) if whenever all experts have single-peaked preferences and all submit their peaks to be aggregated by $g$, no expert can obtain a strictly better alternative by reporting a fake peak. 
\end{definition}
 
 It can be proved that one-SP is equivalent to the following property called \textbf{uncompromisingness} in \cite{BJ83}. A voting rule $g$ is one-SP iff for all experts $i\in N$ and for all peak profiles $\textbf{r}\in [0,1]^N$:
\[
    r_i < g(\textbf{r}) \Rightarrow g(\textbf{r}) \leq g(\textbf{r}_{-i}(r_i'))
\]
and 
\[
    r_i > g(\textbf{r}) \Rightarrow g(\textbf{r}) \geq g(\textbf{r}_{-i}(r_i')).\]

Our Level-SP definition below is a natural extension of uncompromisingness when the input space is the set of cumulative distributions. We will show that it implies "classical" strategy\-proofness for a large class of single-peaked preferences in the cumulative space.

One may be interested in voting rules such that whenever all experts agree on a peak, society chooses that peak. This axiom is called unanimity. We formulate unanimity for a general input space $X$. We will use it for our own characterizations.
\begin{axiom}[Unanimity]
$h: X^N\rightarrow X$ is \textbf{unanimous} if $\forall x\in X$ we have $h(x,x,\dots,x) =x$.
\end{axiom}

Now we give two well-known characterizations of Moulin \cite{M1980} that we need in the next section.

\begin{lemma}\label{moulin g}[Moulin's max-min formula \cite{M1980}]
A \textbf{voting rule} $g : [0,1]^N \rightarrow [0,1]$ is one-SP iff for each coalition of players $S \in 2^N$, there exists a unique value $\beta_S\in [0,1]$ called a \textbf{phantom voter} s.t.\\
\begin{center}
$S \subseteq S'$ implies $\beta_S \leq \beta_{S'}$ and
\end{center}
\[\forall \textbf{r} \in [0,1]^N, g(\textbf{r}) = \max_{S \subseteq N}  \min{(\beta_S, \min_{i\in S} r_i )}.\]

Moreover, the method is unanimous iff $\beta_\emptyset=0$ and $\beta_N=1$.
\end{lemma}

Moulin's most popular ``median'' formula was established in the anonymous case when the experts are treated equally by the rule. We formulate this axiom for any input space $X$ since we will use it later.

\begin{axiom}[Anonymity]
$h: X^N\rightarrow X$ is \textbf{anonymous} if $\forall \textbf{x} \in X^N$ and all permutation $\sigma$ over $N$:
$$h(x_{\sigma(1)},\dots,x_{\sigma(n)} )= h(\textbf{x}).$$
\end{axiom}

\begin{lemma}\label{moulin a}[Moulin's median formula \cite{M1980}]
A voting rule $g : [0,1]^N \rightarrow [0,1]$ is one-SP and anonymous iff there exist $n+1$ \textbf{phantom voters} $\alpha_0\leq\dots\leq \alpha_{n}$ in [0,1] such that:
\[\forall \textbf{r} \in [0,1]^N, g(\textbf{r}) = \operatorname{median}(r_1,\dots,r_n,\alpha_0,\dots,\alpha_{n}).\]
Moreover, the method is unanimous iff $\alpha_0=0$ and $\alpha_{n}=1$.
\end{lemma}
 
\begin{remark}
When $g$ is anonymous, the $\alpha_k$ in the median characterization is equal to $\beta_S$ in the max-min characterization whenever the cardinal of $S$ is $k$ (see \cite{MathProgPaper}). Consequently, the term "phantom voter" is therefore not ambiguous.
\end{remark}

\begin{remark}
     Recently \cite{MathProgPaper}, new characterizations have been established for one-SP rules. For example, it was shown that a voting rule $g : [0,1]^N \rightarrow [0,1]$ is one-SP and anonymous iff there is a weakly increasing function $\gamma:[0,1] \rightarrow [0,1]$ (called the grading curve) such that:
\[\forall \vv{r}, g(\vv{r}) := \sup \left\{ y \middle| \gamma\left(\dfrac{\#\{ r_i \geq y\}}{n}\right) \geq y \right\},\] 
where $\dfrac{\#\{ r_i \geq y\}}{n}$ denotes the number of voters $i\in N$ whose input grade $r_i$ is higher or equal to $y$.\\
In the sequel, we will extend this characterization to the probability aggregation context when voters can have different weights.
%$\gamma$ is called the grading curve. The curve characterization extends to the non anonymous case. 
\end{remark}

\subsection{Definition of level-strategyproofness (Level-SP).}

This subsection describes our probability aggregation model and the new notion of Level-SP. The next subsection links it to ``classical'' strategy\-proofness. 

Society, a decision-maker, or the regulator wants to estimate the probability of a random variable $X$ that ranges over a linearly ordered set of outcomes $\Lambda$ that we identify with a Borel subset of the real line $\mathds{R}$.\footnote{Borel sets are obtained from  countable union, countable intersection, and relative complement of the intervals.} In practical applications, $\Lambda$ is finite or an interval.

 Let $\mathcal{P}$ denote the set of Borel probability distributions over $\Lambda$. To construct society's estimation, each expert $i\in N=\{1,\dots,n\}$ is asked to submit their subjective probability distribution\footnote{A probability distribution is a $\sigma$-additive positive measure over the Borel algebra with total mass equals to 1.} estimation $p_i \in \mathcal{P}$. Our objective is to design a Probability Aggregation Function (PAF) $\psi : \mathcal{P}^N \rightarrow \mathcal{P}$ satisfying some desirable properties. As some experts may be strategically behaving, we wish $\psi$ to be Incentive Compatible (IC). Our IC notion is called Level-SP (level-strategy\-proof). It implies honest reporting when society's decision is based on a threshold. That is to say, the decision is made if the probability of having reached that threshold is above a certain level. Practical applications include situations that are represented with tipping points (e.g. climate change, the degree of a volcanic eruption damaging impact). 

\begin{definition}[Level Events]
We define for any alternative $a\in \Lambda$, the \textbf{level event} $\mathcal{E}(a)$ as ``the threshold $a$ has not been crossed'', e.g. 
\[\mathcal{E}(a)=\{x\in \Lambda: x\leq a\}\]
\end{definition}

The next definition says that a PAF is IC with respect to level events if, whatever threshold $a\in \Lambda$ was chosen, no misreporting expert can orientate society's probability for the level event $\mathcal{E}(a)$ towards their wishes.

\begin{axiom}[Level-SP]
A PAF $\psi : \mathcal{P}^N \rightarrow \mathcal{P}$ is \textbf{level-strategy\-proof} if, for every expert $i\in N$, for every input profile $\textbf{p}\in \mathcal{P}^N$, for every potential deviation $p'_i \in \mathcal{P}$ of expert $i$ and for every threshold $a\in \Lambda$:
\[
    p_i(\mathcal{E}(a)) < \psi(\textbf{p})(\mathcal{E}(a))  \Rightarrow \psi(\textbf{p})(\mathcal{E}(a)) \leq \psi(\textbf{p}_{-i}(p'_i))(\mathcal{E}(a))
\]
and 
\[
    p_i(\mathcal{E}(a)) > \psi(\textbf{p})(\mathcal{E}(a))  \Rightarrow \psi(\textbf{p})(\mathcal{E}(a)) \geq \psi(\textbf{p}_{-i}(p'_i))(\mathcal{E}(a)).
\]
\end{axiom}

This axiom is the uncompromisingness property seen above to be satisfied for all level events.

Let us formulate this in a more practical way. Let $\mathcal{C}$ be the set of \textbf{cumulative distribution functions} (CDF) over $\Lambda$ and, let $\pi: \mathcal{P}\rightarrow \mathcal{C}$ be the mapping  that transforms a probability distribution $p$ over $\Lambda$ into its CDF $P$:
\begin{equation}\label{def d une cdf}
\forall p \in \mathcal{P}, \forall a \in \Lambda, P(a) =\pi(p)(a)= p(\mathcal{E}(a))=\int_{x\leq a} \mathrm{d}p(x). 
\end{equation}

A PAF $\psi:\mathcal{P}^N\rightarrow\mathcal{P}$ is associated with a unique \textbf{cumulative aggregation function (CAF)} $\Psi : \mathcal{C}^N \rightarrow \mathcal{C}$ that takes the CDFs $\{P_i=\pi(p_i)\}_{i\in N}$ of experts as inputs and returns the CDF of $\psi(\textbf{p})$ as output.

\[\Psi(\pi(p_1),\dots,\pi(p_n))=\pi(\psi(p_1,\dots,p_n)).\]

With this notation, stating that the PAF $\psi:\mathcal{P}^N\rightarrow\mathcal{P}$ is Level-SP is equivalent to stating that the associated CAF $\Psi:\mathcal{C}^N\rightarrow\mathcal{C}$ satisfies: $\forall a\in \Lambda$, every $\textbf{P}\in \mathcal{C}^N$, $\textbf{P'} \in \mathcal{C}^N$ and all $i\in N$:
\[
    P_i(a) < \Psi(\textbf{P})(a)  \Rightarrow \Psi(\textbf{P})(a) \leq \Psi(\textbf{P}_{-i}(P'_i))(a)
\]
and 
\[
    P_i(a) > \Psi(\textbf{P})(a)  \Rightarrow \Psi(P)(a) \geq \Psi(\textbf{P}_{-i}(P'_i))(a).
\]
This implies that $\forall a\in \Lambda$, $\textbf{P}\in \mathcal{C}^N$, $\textbf{P'} \in \mathcal{C}^N$ and $i\in N$:
\begin{equation}\label{inegalite}
| \Psi(\textbf{P}_{-i}(P'_i))(a) - P_i(a) | \geq   | \Psi(\textbf{P})(a) - P_i(a) |.
\end{equation}

\begin{remark}
Level-SP is better formulated in the CAF space, and one may wonder why we don't just use the CAF model. The reason is: all the remaining axioms (namely certainty preservation, plausibility preservation, and proportionality) are more naturally formulated in the PAF model.  
  
\end{remark}

\subsection{The rich domain of utility functions related to Level-SP.}

As the regulator does not usually know exactly the preferences of the experts, it is desirable to have truthful reporting property for a large family of reasonable utility functions. This section shows that Level-SP is a robust IC concept as it implies ``classical'' strategy\-proofness for a large and natural class of single-peaked utility functions. 

For each probability measure $\nu$ on $\Lambda$ and each positive real $r\in \mathds{R}^*_+$, we can define an $L_r$-distance on $\mathcal{C}$ as follows:
$\| P-Q \|_{L_r}=[ \int_\Lambda | P(a)-Q(a)|^r d\nu (a) ]^{1/r} $. 

If $\psi$ verifies Level-SP, it also satisfies, $\forall i\in N$, $\forall \textbf{P}\in \mathcal{C}^N$ and any $ P'_i \in \mathcal{C}$:

\begin{align*}
    \|\Psi(\textbf{P}_{-i}(P'_i)) - P_i\|_{L_r} &= \left[\int_\Lambda | \Psi(\textbf{P}_{-i}(P'_i))(a) - P_i(a) |^r \mathrm{d}\nu (a) \right]^{1/r} \\
    &\geq \left[\int_\Lambda | \Psi(\textbf{P})(a) - P_i(a) |^r \mathrm{d}\nu (a) \right]^{1/r} \\
    &= \|\Psi(\textbf{P}) - P_i\|_{L_r}
\end{align*}

where the inequality follows from (\ref{inegalite}). Consequently, if the utility function $u_i$ of expert $i\in N$ is single-peaked and is measured by the use of some distance $L_{r}$ on $\mathcal{C}$ to its peak, e.g.:
\[ u_i(p)=-\| \pi(\psi(\textbf{p})) - \pi(p_i)\|_{L_{r}}=-\| \Psi(\textbf{P}) - P_i\|_{L_{r}}\]
then it is an optimal strategy for expert $i$ with the utility function just above to vote honestly.

\section{The characterization of all Level-SP methods.}\label{characterization}

In this section, we will characterize all Level-SP methods. They consist of aggregating the cumulatives instead of the probabilities and then using Moulin's formulae by replacing the phantom voters with weakly increasing functions satisfying regularity conditions.

As previously stated in \ref{def d une cdf}, the cumulative distribution function (CDF) $P=\pi(p) : \Lambda \rightarrow [0,1]$ associated with a probability distribution $p$ on $\mathds{R}$ is given by the formulae $P(a)=\int_{x\leq a} dp(x)$. The following lemma is well-known. It helps understand the regularity conditions in the next theorem.

\begin{lemma}\label{ lemma charac distrib}
A function $P : \Lambda \rightarrow [0,1]$ is a cumulative distribution iff
\begin{itemize}
    \item $P$ is weakly increasing and right continuous;
    
    \item if $\sup \Lambda \not \in \Lambda$ then $\lim_{a \rightarrow\sup\Lambda}P(a) = 1$, otherwise $P(\sup \Lambda)=1$;
    \item if $\inf \Lambda \not \in \Lambda$ then $\lim_{a \rightarrow\inf\Lambda}P(a) = 0$. 
\end{itemize}

\end{lemma}

For example, if $\Lambda =\mathds{R} $ then $\inf \Lambda = - \infty \not \in \Lambda$ and $\sup \Lambda = + \infty \not \in \Lambda$. 

With this lemma and Moulin max-min formula, we can prove the following characterization.

\begin{theorem}[Level-SP: Max-min formula]\label{charac general}
A PAF $\psi : \mathcal{P}^N \rightarrow \mathcal{P}$ is Level-SP iff there exists for every $S \subseteq N$ a weakly increasing, right continuous functions $f_S : \Lambda \rightarrow [0,1]$ that verify the following properties:
\begin{itemize}
    \item (1) $\forall S \subseteq S'$ and $ \forall a\in \Lambda$,  $f_S(a) \leq f_{S'}(a)$;
    \item (2) if $\sup \Lambda \not \in \Lambda$, then $\lim_{a \rightarrow\sup\Lambda}f_N(a) = 1$, otherwise $f_N(\sup\Lambda) = 1$ ;
    \item (3) if $\inf \Lambda \not \in \Lambda$ then $\lim_{a \rightarrow\inf\Lambda}f_\emptyset(a) = 0$;
    \item (4) the CAF $\Psi : \mathcal{C}^N \rightarrow \mathcal{C}$ associated with $\psi$ is given by the formula:
\[\forall a \in \Lambda, \Psi(\textbf{P})(a) = \pi \circ \psi(\textbf{p})(a)  =\max_{S \subseteq N}  \min{(f_S(a), \min_{i\in S} P_i(a))}. \]
\end{itemize}
 Moreover, $\psi$ is unanimous iff for every $a\in \Lambda$  $f_{\emptyset}(a)=0$ and $f_{N}(a)=1$ (in which case, (2) and (3) are automatically satisfied).\\
 \textbf{We will call the $\{a\rightarrow f_S(a)\}_{S\in 2^N}$ above the phantom functions associated with $\psi$}. 
\end{theorem}

The technical conditions (2) and (3) are necessary for $\Psi$ to output a cumulative distribution. Conditions (1) and (4) are derived from to the Moulin max-min formula. 

\proof{Sketch of proof}
In order to prove this theorem, we will first prove that Level-SP implies that $\forall a\in \Lambda$, there is a one-SP function $g_a:[0,1]^N \rightarrow [0,1]$ such that:
\begin{equation}\label{* ref}
    \forall \textbf{P}, \Psi(\textbf{P})(a) = g_a(P_1(a),\dots,P_n(a)).
\end{equation}
The rest is a direct consequence of Lemma \ref{moulin g} and Lemma \ref{ lemma charac distrib}.

Let us prove the existence of such $\{g_a\}_{a\in \Lambda}$ by \emph{reductio ad absurdum}.
Suppose that $\Psi$ verifies Level-SP but that there exists an alternative $a\in \Lambda$ such that no voting rule $g_a$ (one-SP or not) is such that (\ref{* ref}) is satisfied. Then there must exist two CDF profiles $\textbf{P}$ and $\textbf{Q}$ such that $\forall i\in N$, $P_i(a) = Q_i(a)$ and $\Psi(\textbf{P})(a) \neq \Psi(\textbf{Q})(a)$.  We will show that switching experts inputs, one by one, from $P_i$ to $Q_i$, one at a time, does not change the output. This will result in a contradiction with the assumption $\Psi(\textbf{P})(a) \neq \Psi(\textbf{Q})(a)$. Wlog, suppose $\Psi(\textbf{P})(a) < \Psi(\textbf{Q})(a)$. 

Here is the proof for switching expert 1's opinion when $P_1(a) < \Psi(\textbf{P})(a)$ (the proof for $P_1(a) > \Psi(\textbf{P})(a))$ is symmetrical):
\begin{itemize}
    \item By Level-SP, we have $\Psi(\textbf{P})(a) \leq \Psi(\textbf{P}_{-1}(Q_1))(a)$.
    \item Since $Q_1(a)=P_1(a)$ we therefore have $Q_1(a) < \Psi(\textbf{P}_{-1}(Q_1))(a)$.
    \item By Level-SP, we therefore have $\Psi(\textbf{P})(a) \geq \Psi(\textbf{P}_{-1}(Q_1))(a)$. It follows that $\Psi(\textbf{P})(a) = \Psi(\textbf{P}_{-1}(Q_1))(a)$.
\end{itemize}

If remains to consider the case $\Psi(\textbf{P})(a) = P_1(a) = Q_1(a) \neq \Psi(\textbf{P}_{-1}(Q_1)(a))$. This would contradict Level-SP since by switching expert $1$'s input from $Q_1$ to $P_1$ the output becomes the value $Q_1$.

 Consequently, when we switch expert $1$'s input from $P_1$ to $Q_1$ we do not change the output. The same proof is then repeated for all other experts. It follows that our assumption was wrong: the family of voting rules $\{g_a\}_{a\in \Lambda}$ where (*) is satisfied exists. Due to the definition of Level-SP, each $g_a$ must be one-SP. As such, from Lemma \ref{moulin g}, there exist phantoms $(\beta_a^S)_{S\subset N}$ associated with each $g_a$. We define the $f_S$ in the theorem as follows $f_S : a \rightarrow \beta_a^S$. The properties that $f_S$ must verify are simply those needed so that the outcome of $\Psi$ is a cumulative distribution (see lemma \ref{ lemma charac distrib}).

A more detailed and direct proof (not by contradiction) can be found in appendix \ref{proof of the theorem general}.

\endproof

We now provide a similar characterization for the anonymous case.

\begin{theorem}[Level-SP: the Median formula]\label{anonymous charac}
A PAF $\psi : \mathcal{P}^N \rightarrow \mathcal{P}$ is Level-SP and anonymous iff there exists $n+1$ weakly increasing, right continuous functions $f_k : \Lambda \rightarrow [0,1]$ that verify the following properties:
\begin{itemize}
    \item for all $0\leq k \leq n-1$ we have $f_k \leq f_{k+1}$;
    \item 
    if $\sup \Lambda \not \in \Lambda$, then $\lim_{a \rightarrow\sup\Lambda}f_n(a) = 1$ otherwise $f_n(\sup\Lambda) = 1$ 
    \item if $\inf \Lambda \not \in \Lambda$ then $\lim_{a \rightarrow\inf\Lambda}f_0(a) = 0$;
    \item  the CAF $\Psi : \mathcal{C}^N \rightarrow \mathcal{C}$ associated with $\psi$ is given by the formula:
\[\forall a \in \Lambda, \Psi(\textbf{P})(a) =  \operatorname{median}\left(P_1(a),\dots,P_n(a),f_0(a),\dots,f_{n}(a)\right).\]
\end{itemize}
 Moreover $\psi$ is unanimous iff for all $a\in \Lambda$ $f_0(a)=0$ and  $f_{n}(a)=1$.\\ 
 \textbf{We will call the $\{f_k\}_{k\in \{1,\dots,n+1\}}$ the phantom functions associated with $\psi$}. 

\end{theorem}

\proof{Proof} Essentially the same as in Theorem \ref{charac general} except we use Lemma \ref{moulin a} instead of Lemma \ref{moulin g} (see appendix \ref{proof of the theorem general}). 

\endproof

\begin{remark}\label{remark consistent phantoms}
  As in Moulin, it can be shown that, under anonymity, the phantom function $f_S$ in the max-min formula is equal to $f_k$ in the median formula where $k=\#S$, the cardinal of $S$.
\end{remark}

\proof{Proof}
The proof can be found in the appendix \ref{proof consistent phantoms}. 
\endproof
%{\color{green}
\begin{remark}
    Alternatively, if we use the grading curve characterization in \cite{MathProgPaper}, we conclude that a PAF $\psi : \mathcal{P}^N \rightarrow \mathcal{P}$ is Level-SP and anonymous iff there exists a family $(\gamma_a)_{a_\in \Lambda}$ of grading curves\footnote{Weakly increasing functions $\gamma_a:[0,1] \rightarrow [0,1]$ satisfying some regularity conditions w.r.t. to $a\in \Lambda$ such as right continuity, weak monotony and some boundary conditions.} such that:
\[\forall a \in \Lambda, \Psi(\textbf{P})(a) := \sup \left\{ y \middle| \gamma_a\left(\dfrac{\#\{ P_i(a) \geq y\}}{n}\right) \geq y \right\}.\]
\end{remark}
%}

\begin{figure}
   \centering
\begin{subfigure}{0.48\textwidth}
\centering
\scalebox{0.8}{
\begin{tikzpicture}
\begin{axis}[
    legend pos=south east,
    axis lines = left,
    xlabel = $x$,
    ylabel = $P_i(x)$,
]

\addplot [
    domain=0:5, 
    samples=100, 
    color=blue,
    ]
    {1-exp(-2*x)};
\addlegendentry{$P_1=1-exp(-2x)$}

\addplot [
    domain=0:5, 
    samples=100, 
    color=red,
]
{1-exp(-x)};
\addlegendentry{$P_2=1-exp(-x)$}

\addplot [
    domain=0:5, 
    samples=100, 
    color=brown,
    ]
    {1-exp(-0.5*x)};
\addlegendentry{$P_3=1-exp(-0.5x)$}

\addplot[
    domain=0:5, 
    samples=100, 
    style = thick,
    color=black]{0.25+ 0.1*x};
\addlegendentry{Phantoms $F_2$ and $F_3$}

\addplot[
    domain=0:5, 
    samples=100, 
    style = thick,
    color=black]{min(0.5+ 0.05*x^2,0.8)};

\end{axis}
\end{tikzpicture}}
\end{subfigure}
\hfill
\begin{subfigure}{0.48\textwidth}
\centering
\scalebox{0.8}{
\begin{tikzpicture}
\begin{axis}[
    legend pos=south east,
    axis lines = left,
    xlabel = $x$,
    ylabel = $P_i(x)$,
]

\addplot [
    domain=0:5, 
    samples=100, 
    color=blue,
    ]
    {1-exp(-2*x)};
\addlegendentry{$P_1=1-exp(-2x)$}

\addplot [
    domain=0:5, 
    samples=100, 
    color=red,
]
{1-exp(-x)};
\addlegendentry{$P_2=1-exp(-x)$}

\addplot [
    domain=0:5, 
    samples=100, 
    color=brown,
    ]
    {1-exp(-0.5*x)};
\addlegendentry{$P_3=1-exp(-0.5x)$}

\addplot[
    domain=0:5, 
    samples=100, 
    style = thick,
    color=black]{0.25+ 0.1*x};
\addlegendentry{Phantoms $F_2$ and $F_3$}

\addplot[
    domain=0:0.16, 
    samples=30, 
    style={ultra thick},
    color=green]{1-exp(-2*x)};
\addlegendentry{$med(P_1,P_2,P_3,F_2,F_3)$}

\addplot[
    domain=0:5, 
    samples=100, 
    style = thick,
    color=black]{min(0.5+ 0.05*x^2,0.8)};

%sol

\addplot[
    domain=0.16:0.315, 
    samples=30, 
    style={ultra thick},
    color=green]{0.25+ 0.1*x};
    
\addplot[
    domain=0.315:0.79, 
    samples=30, 
    style={ultra thick},
    color=green]{1-exp(-x)};
    
\addplot[
    domain=0.79:3.2, 
    samples=30, 
    style={ultra thick},
    color=green]{min(0.5+ 0.05*x^2,0.8)};

\addplot[
    domain=3.2:5, 
    samples=30, 
    style={ultra thick},
    color=green]{1-exp(-0.5*x)};

\end{axis}
\end{tikzpicture}}
\end{subfigure}
\caption{\small A unanimous and anonymous Level-SP rule for $\Lambda=\textbf{R}_+$. In left hand is drawn the cumulative functions $P_1$, $P_2$ and $P_3$ of 3 voters. In black is drawn two the phantom functions $F_2$ and $F_3$ ($F_1=0$ and $F4=1$ are not drawn). On the right hand, for each level $x$ in the x-axe, the outcome in green is determined by the median formula: e.g. $median(P_1(x), P_2(x), P_3(x), F_2(x), F_3(x)$).}
\label{anonymous ex}
\end{figure}
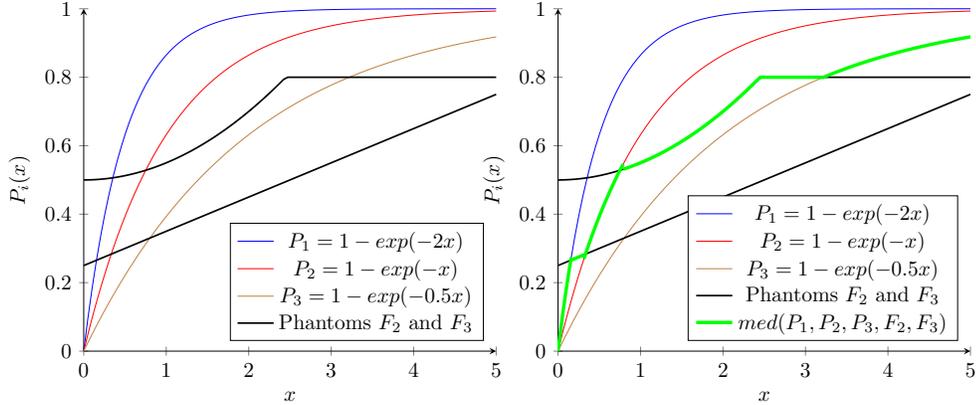

Figure \ref{anonymous ex} provides an example with $\Lambda=\mathds{R}_+$ and $3$ experts. The next sections refine the characterizations by adding desirable (and classical) axioms. Some combination of the axioms will single out only one method, or lead to an impossibility.

\section{Characterizing certainty preserving Level-SP methods.}\label{section certainty}

It is desirable, and somehow incontestable, to wish that when all experts agree that an event is certain to happen (or not to happen) then the aggregation of their input probabilities reflects that fact.

\begin{axiom}[Certainty preservation]
A PAF $\psi$ \textbf{preserves certainty} if for any probability profile $\textbf{p}$ and all events $A \subseteq \Lambda$ Borel measurable:
\begin{align*}
    & ( p_i(A)=1  \ \ \forall i\in N \Rightarrow \psi(\textbf{p})(A)=1),  \\  \text{or equivalently}, \  & ( p_i(A)=0 \ \  \forall i\in N \Rightarrow \psi(\textbf{p})(A)=0).
\end{align*}
\end{axiom}

 In other words, if an event is judged impossible (resp. certain) by all experts then it is judged impossible (resp. certain) by the aggregation function. This is a fairly standard axiom in the literature sometimes called the \textit{zero preservation property} \cite{MartiniS} or \textit{consensus preservation} \cite{DL2017}.

\begin{proposition}[Certainty preservation characterization]\label{certainty pres}
A Level-SP PAF $\psi$ is certainty preserving iff it is unanimous and its associated phantom functions $f_S$ are constants.
\end{proposition}

\proof{Sketch of proof} (Sketch of $\Rightarrow:$) For any interval $I=[a,b] \cap \Lambda$ and $S$. Select a profile $\textbf{P}$ such that all experts in $S$ agree that $\mathcal{E}(a)$ is certain almost surely and all other experts agree that $\mathcal{E}(b)$ is impossible almost surely. It follows that $I$ is impossible almost surely so the outcome will be constant over $I$ and that the outcome is equal to $f_S$ over $I$. Therefore $f_S$ is constant over $I$.  It follows that all $f_S$ are constant over $\Lambda$. For the complete proof, see appendix \ref{certainty appendix}. 

\endproof

In other words a Level-SP $\psi$ is certainty preserving if and only if there exists $2^n$ phantom constant values $\beta_S \in [0,1]$ such that $\beta_\emptyset=0$, $\beta_N=1$ and $\Psi : \mathcal{C}^N \rightarrow \mathcal{C}$ the CAF associated with $\psi$ verifies the following:
\[\forall a \in \Lambda, \Psi(\textbf{P})(a) =\max_{S \subseteq N}  \min{ (\beta_S, \min_{i\in S} P_i(a))}. \]

In the anonymous case the characterization simplifies to:
\[\forall a \in \Lambda, \Psi(\textbf{P})(a) =\operatorname{median}(P_1(a),\dots,P_n(a),\alpha_1,\dots,\alpha_{n-1}). \]

Where $\alpha_k=\beta_S$ whenever $\#S=k$. When $\psi$ is Level-SP and certainty preserving, the one-SP function $g:\Lambda^n\rightarrow \Lambda$ such that: \[\forall a \in \Lambda, \Psi(\textbf{P})(a)= g(\textbf{P}(a))\] is called in the sequel \textbf{the one-SP rule associated with $\Psi$}. In the general case, $g(r_1,...,r_n)=\max_{S \subseteq N}  \min{ (\beta_S, \min_{i\in S} r_i)}$ and in the anonymous case, $g(r_1,...,r_n)=\operatorname{median}(r_1,\dots,r_n,\alpha_1,\dots,\alpha_{n-1}).$

%\begin{remark}\label{remark curve}
 %   If we use the grading curve characterization in \cite{MathProgPaper}, we conclude that a PAF $\psi : \mathcal{P}^N \rightarrow \mathcal{P}$ is Level-SP,  anonymous and certainty preserving iff it is unanimous and there is a grading curve $\gamma:[0,1] \rightarrow [0,1]$ such that :
%\[\forall a \in \Lambda, \Psi(\textbf{P})(a) := \sup \left\{ y | \dfrac{\#\{ P_i(a) \geq y\}}{n} \geq \gamma(y) \right\}.\] 
%\end{remark}

%\begin{corollary}\label{corollary curve}
 %       A Level-strategyproof and certainty preserving WPAF $\psi_{\text{W}} :\mathcal{P}^N \rightarrow \mathcal{P}$ verifies weight proportionality and additivity iff there exists $\gamma$, a grading curve\footnote{Increasing functions $\gamma:[0,1] \rightarrow [0,1]$ with $\gamma(y) > 0$ for $y > 0$} such that $\forall a\in \Lambda$ and $\forall M \subset W$ finite, the cumulative $\Psi^M_{\text{W}}$ of $\psi_{\text{W}} $ satisfies

%\[\Psi^M_{\text{W}}(\textbf{P})(a) := \sup \left\{ y \middle| \dfrac{\sum_{i\in M: P_i(a) \geq y}w_i}{\sum_{i \in M}w_i} \geq \gamma(y) \right\}.\] 
 
%\end{corollary}

\subsection{Relationship between certainty preservation and the independence axiom.}

\begin{axiom}[Level Independence]
A PAF $\psi$ satisfies \textbf{level independence} if there is a surjective voting rule $g:[0,1]^N\rightarrow[0,1]$ such that:
\[
\forall a \in \Lambda, \forall \textbf{P}\in \mathcal{C}^N, \Psi(\textbf{P})(a)=g(P_1(a),...,P_N(a)).
\]
\end{axiom}

Hence, a Level-SP PAF $\psi$ is certainty preserving iff it satisfies level independence. Level-independence is in fact, a weak version of a classical axiom called \textit{independence}. 

\begin{axiom}[Independence]
The PAF $\psi$ satisfies \textbf{independence} if there is $g:[0,1]^N\rightarrow [0,1]$ such that $\psi(p_1,...,p_n) (A)=g(p_1(A),...,p_n(A))$ for every input $\textbf{p}=(p_1,...,p_n)$ and every Borel measurable subset $A\subset \Lambda$.
\end{axiom}

Hence, level-independence is independence when the test is restricted to only Borel sets that are level sets (e.g. $A=\{x\in \Lambda: x\leq a \}$, $a\in \Lambda$). Unfortunately, independence is known to be a very strong axiom as, alone, it implies that $\psi$ must be a weighted average of the expert's inputs \cite{Stone1961, McConway1981, BP,CGV, DL2017}, which is not level-strategyproof unless it is dictatorial. Consequently:

\begin{theorem}[Impossibility]
Level-SP and independent PAFs are the dictatorials.
\end{theorem}

\section{Charaterizing plausibility preservating Level-SP methods.}\label{section plausibility}

In this section, we characterize the PAFs that verify that if all experts agree that an outcome may happen with positive probability then the PAF preserves this property. 

\begin{axiom}[Plausibility preservation]
A PAF $\psi :\mathcal{P}^N \rightarrow \mathcal{P}$ verifies plausibility preservation if for any input profile $\textbf{p}\in \mathcal{P}^N$ and any possible interval $I= [a,b] \cap \Lambda$:
\[ p_i(I) > 0 \ \ \forall i \in N \implies \psi(\textbf{p})(I) > 0. \]
\end{axiom}

Plausibility preservation implies that some monotonicity property is satisfied by the phantom functions in all intervals where their values are not 0 or 1. 

\begin{proposition}[Plausibility preservation characterization]\label{plaus charac}
A Level-SP PAF $\psi : \mathcal{P}^N \rightarrow \mathcal{P}$ is plausibility preserving iff: 
\begin{enumerate}
    \item Each associated phantom function $f_S$ is strictly increasing on the interval where its value is not in $\{0,1\}$;
    \item $\forall a < \sup\Lambda, f_\emptyset(a) < 1$ and $\forall a\in \Lambda, f_N(a) > 0$.
\end{enumerate}
\end{proposition}

\proof{ Sketch of proof}
(Sketch of $\Rightarrow:$) For any interval $I=[a,b] \cap \Lambda$, any $S \subseteq N$ and any $0<\epsilon <0.5$. Select a profile $\textbf{P}$ such that all experts in $i \in S$ agree that $p_i(\mathcal{E}(a)) = 1-\epsilon$, $p_i(\mathcal{E}(b)) = 1$ and all other experts $j$ agree that $p_j(\mathcal{E}(a)) = 0$ and $p_j(\mathcal{E}(b)) = \epsilon$. It follows that all experts agree that $p_i(I)>0$ Therefore $\Psi(\textbf{P})(b) - \Psi(\textbf{P})(a) >0$. If $f_S(I) =x \in ]0,1[$, then by choosing $\epsilon$ such that $\epsilon <x <1-\epsilon$ we have $\Psi(\textbf{P})(b)- \Psi(\textbf{P})(a) =0$. This is absurd since the outcome should have determined that $I$ is possible, therefore we contradicted $f_S(I) = x \in ]0,1[$. The rest can be deduced from this. 

A more detailed proof can be found in the appendix \ref{plausibility preservation}. 

 \endproof

Figure \ref{cert ex} represents a certainty preserving Level-SP PAF because the phantom functions $F_0=0, F_1=0.25, F_2=0.6, F_3=0.6$ are constant (as in Figure \ref{anonymous ex}, we didn't draw $F_0$ and $F_4$, for simplicity and also because the median is the same with or without them). However, it is not plausibility preserving because $F_3=0.6$ is not in $\{0,1\}$ in the interval $[0,1]$ but is not strictly increasing (is constant). As such, the green function (the CDF of society) is constant on the interval $[1,2]$ implying that the probability of $[1,2]$ of society is 0 although all experts give a positive probability to $[1,2]$.

Figure \ref{plaus ex} represents a plausibility preserving Level-SP PAF because two phantom functions (not drawn) $F_0=0$ and $F_4=1$ have values in $\{0,1\}$ and the two others $F_2$ and $F_3$ are strictly increasing. Observe that $F_3$ is discontinuous at its left for $x=2$ (which is allowed for phantom functions). Since $F_2$ is not a constant function, the PAF is not certainty preserving.

%certainty and plausibility graphs
\begin{figure}
\centering
\begin{minipage}{0.48\textwidth}
\centering
\scalebox{0.8}{
\begin{tikzpicture}{scale=0.8}
\begin{axis}[
    legend pos=south east,
    axis lines = left,
    xlabel = $x$,
    ylabel = $P_i(x)$,
]

\addplot [
    domain=0:5, 
    samples=100, 
    color=blue,
    ]
    {1-exp(-2*x)};
\addlegendentry{$P_1=1-exp(-2x)$}

\addplot [
    domain=0:5, 
    samples=100, 
    color=red,
]
{1-exp(-x)};
\addlegendentry{$P_2=1-exp(-x)$}

\addplot [
    domain=0:5, 
    samples=100, 
    color=brown,
    ]
    {1-exp(-0.5*x)};
\addlegendentry{$P_3=1-exp(-0.5x)$}

\addplot[
    domain=0:5, 
    samples=100, 
    color=black]{0.25};
\addlegendentry{Phantoms $F_2$ and $F_3$}
    
\addplot[
    domain=0:0.15, 
    samples=30, 
    style={ultra thick},
    color=green]{1-exp(-2*x)};
\addlegendentry{$med(P_1,P_2,P_3,F_2,F_3)$}

\addplot[
    domain=0:5, 
    samples=100, 
    color=black]{0.6};

\addplot[
    domain=0.15:0.3, 
    samples=30, 
    style={ultra thick},
    color=green]{0.25};
    
\addplot[
    domain=0.3:0.9, 
    samples=30, 
    style={ultra thick},
    color=green]{1-exp(-x)};
    
\addplot[
    domain=0.9:1.8, 
    samples=30, 
    style={ultra thick},
    color=green]{0.6};

\addplot[
    domain=1.8:5, 
    samples=30, 
    style={ultra thick},
    color=green]{1-exp(-0.5*x)};

\end{axis}
\end{tikzpicture}}
\captionof{figure}{A certainty preserving but not plausibility preserving, unanimous and anonymous method.}
\label{cert ex}
\end{minipage}
\hfill
\begin{minipage}{0.48\textwidth}
\centering
\scalebox{0.8}{
\begin{tikzpicture}
\begin{axis}[
    legend pos=south east,
    axis lines = left,
    xlabel = $x$,
    ylabel = $P_i(x)$,
]

\addplot [
    domain=0:5, 
    samples=100, 
    color=blue,
    ]
    {1-exp(-2*x)};
\addlegendentry{$P_1=1-exp(-2x)$}

\addplot [
    domain=0:5, 
    samples=100, 
    color=red,
]
{1-exp(-x)};
\addlegendentry{$P_2=1-exp(-x)$}

\addplot [
    domain=0:5, 
    samples=100, 
    color=brown,
    ]
    {1-exp(-0.5*x)};
\addlegendentry{$P_3=1-exp(-0.5x)$}

\addplot[
    domain=0:5, 
    samples=100, 
    color=black]{0.25+ 0.1*x};
\addlegendentry{Phantoms $F_2$ and $F_3$}

\addplot[
    domain=0:0.16, 
    samples=30, 
    style={ultra thick},
    color=green]{1-exp(-2*x)};
\addlegendentry{$med(P_1,P_2,P_3,F_2,F_3)$}

\addplot[
    domain=0:2, 
    samples=50, 
    color=black]{0.6+ 0.05*x};

\addplot[
    domain=2:5, 
    samples=50, 
    color=black]{0.75+ 0.025*x};

\addplot[
    domain=0.16:0.31, 
    samples=30, 
    style={ultra thick},
    color=green]{0.25+ 0.1*x};
    
\addplot[
    domain=0.31:1.05, 
    samples=30, 
    style={ultra thick},
    color=green]{1-exp(-x)};
    
\addplot[
    domain=1.05:2, 
    samples=30, 
    style={ultra thick},
    color=green]{0.6+ 0.05*x};
    
\addplot[
    domain=2:3.6, 
    samples=30, 
    style={ultra thick},
    color=green]{0.75+0.025*x};

\addplot[
    domain=3.6:5, 
    samples=30, 
    style={ultra thick},
    color=green]{1-exp(-0.5*x)};

\end{axis}
\end{tikzpicture}}
\captionof{figure}{A plausibility preserving but not certainty preserving, unanimous and anonymous method.}
\label{plaus ex}
\end{minipage}
\end{figure}

  It may seem asymmetrical that plausibility preservation is defined by putting a condition on the intervals while certainty preservation is defined by putting conditions on all Borel subsets. First, as the proof shows, if in the definition of certainty preservation, one replaces Borel sets with intervals, we obtain the exact same characterization. However, if we replace intervals with Borel subsets in the definition of plausibility, we reach an impossibility. 

\begin{axiom}[Strong plausibility preservation]
A probability aggregation function $\psi :\mathcal{P}^N \rightarrow \mathcal{P}$ verifies strong plausibility preservation if for any input profiles $\textbf{p}\in \mathcal{P}^N$ and any Borel measurable event $A \subset \Lambda$:
\[ p_i(A) > 0 \ \ \forall i \in N \rightarrow \psi(\textbf{p})(A) > 0; \]
that is, if all experts agree that some Borel event is plausible, so does society. 
\end{axiom}

Unfortunately, extending the plausibility preservation to Borel subsets is a too strong an axiom.

\begin{theorem}[Strong Plausibility Impossibility]\label{ impossibility strong_plau}
When $\Lambda$ is a real interval, the unique PAFs that are Level-SP, unanimous, and strong plausibility preserving are the dictatorials.
\end{theorem}

\proof{ Sketch of proof}
(Sketch of $\Rightarrow$) This is done by reducing to the absurd. We build two disjoint intervals $I_1$ and $I_2$ in $\Lambda$ and a profile $\textbf{P}$, such that we can divide the experts into two groups $S$ and $T$. Experts in $S$ believe that $I_1$ is impossible almost surely and that $I_2$ is possible, experts in $T$ believe that $I_2$ is impossible almost surely and that $I_1$ is possible. In the outcome, we select an expert of $S$ opinion over $I_1$ and of $T$ over $I_2$. By strong plausibility, we should have that the outcome claims $I_1 \cap I_2$ is possible. Therefore we reach a contradiction. See appendix \ref{strong plausibility} for the complete proof. 

\endproof

%{\color{green}

%\begin{theorem}[Strong Plausibility Impossibility 2]\label{Strong Impossibility 2}
%When $\#\Lambda \geq 5$, there are no probability aggregation functions Level-SP, unanimous, anonymous and strong plausibility preserving.
%\end{theorem}

%\begin{theorem}[Strong Plausibility Impossibility 3] \label{Strong Impossibility 3}
%When $\Lambda = \mathbf{Z}$,  the unique probability aggregation functions that are Level-SP, unanimous, and strong plausibility preserving are the dictatorials.
%\end{theorem}

\section{Characterizing the combination of plausibility and certainty preservations.}\label{section combining}

\subsection{The combination of the axioms in the general case.}
Having defined the previous two axioms and how they are characterized in the Level-SP setting, let us combine the two.

\begin{proposition}[Combination 1st Characterization]\label{Combi charac 1}
A Level-SP $\Psi:\mathcal{C}^N \rightarrow \mathcal{C}$ is certainty preserving and plausibility preserving iff the output of its associated one-SP voting rule $g$ is always one of its inputs (e.g. $g(r_1,...,r_n) \subset \{r_1,...,r_n\}$ for every $(r_1,...,r_n)\in [0,1]^N$).  
\end{proposition}

\proof{Sketch of proof} By certainty preservation all phantoms $f_S$ are constant (to some $\beta_S$). By plausibility preservation, all the $\beta_S$ must take their values in $\{0,1\}$. This easily implies the property we want to prove because this implies that the output of $\Psi$ only depends on the ordering of the input. The exact details are found in appendix \ref{combining appendix}. 

\endproof

\begin{remark}
  Hence, a dictatorship is Level-SP, certainty and plausibility preserving.
\end{remark}

This characterization implies the following interesting action on dominated profiles, defined now.

\begin{definition}[Dominated profiles]
A CDF $P$ dominates $Q$ if $P(a) \geq Q(a)$ for all $a\in \Lambda$. A profile $(P_1,..,P_n)$ is dominated, if for any $i \neq j$, or $P_i$ dominates $P_j$ or $P_j$ dominates $P_i$.
\end{definition}

\begin{proposition}[Combination 2nd Characterization]\label{Combi charac 2}
A Level-SP $\psi:\mathcal{P}^N \rightarrow \mathcal{P}$ is certainty and plausibility preserving iff for any dominated profile $\textbf{P}=(P_1,..,P_n)$, $\psi (\textbf{p}) \in \{p_1,...,p_n\}$.
\end{proposition}

\proof{Sketch of proof}
 For a given one-SP voting rule $g$ such that all the phantoms $\beta_S$ are in $\{0,1\}$, the outcome only depends on the ordering of the inputs. When the inputs of a CAF are dominated, the ordering is the same for all levels. As such
for all $a\in \Lambda$, the same expert has their input selected (see appendix \ref{combining appendix} for the complete proof). 

\endproof

Hence, the combination of certainty and plausibility preservation with Level-SP implies that the output probability is one of the inputs whenever the input profile is dominated.

\subsection{The combination of the axioms in the anonymous case.}

When anonymity is added to the combination, we obtain a nice class of PAF that can be generated from a well-known family of one-SP rules, sometimes called the order functions \cite{BL2011}.

\begin{definition}[Order-functions]
An \textbf{order-function} $g^k : [0,1]^n \rightarrow [0,1]$, (where $k$ is in $\{1,\dots,n\}$) is the one-SP voting rule that for a set of $n$ values in $[0,1]$ returns the $k$\textsuperscript{th} smallest value.
\end{definition}

It follows that $g^1(r_1,\dots,r_n)=\operatorname{min}(r_1,\dots,r_n)$, $g^n(r_1,\dots,r_n) = \operatorname{max}(r_1,\dots,r_n)$ and when $n$ is odd \[g^{(n+1)/2} (r_1,\dots,r_n) = \operatorname{median} (r_1,\dots,r_n).\]

\begin{definition}[Order-cumulatives]
For all $k \in \{1,\dots,n\}$, we denote \textbf{$k^{th}$-order-cumulative} $\Psi^k :\mathcal{C}^N \rightarrow \mathcal{C}$ the CAF which is defined by applying the $k^{th}$ order function for each alternative $a\in \Lambda$ in the CDF space:
\[\forall a\in \Lambda, \Psi^k(\textbf{P})(a) := g^k(P_1(a),\dots,P_n(a)).\]
$\psi^k$ denotes the associated PAF that will also be called an order-cumulative.
\end{definition}

Consider a safe-to-dangerous scale $\Lambda$  such as the Richter magnitude scale for earthquakes or the volcanic explosivity index. The min order-cumulative $\Psi^1$ is the most cautious res\-ponse: since for each threshold, we consider the opinion of the most worried expert. On the other hand, the max order-cumulative $\Psi^n$ is the least worried response. 

\begin{theorem}[Order Cumulatives Characterization]
\label{order charac}
The order-cumulatives are the unique Level-SP PAF that are anonymous, certainty preserving, and plausibility preserving. 
\end{theorem}

\proof{ Sketch of proof}
(sketch) $\psi$ is Level-SP therefore their is an associated voting rule $g$. $\psi$ is Level-SP and plausibility preserving therefore the phantoms are equal to $0$ or $1$. As such there is a $k$ such that:
\[\forall a, \Psi(\textbf{P})(a) = med(P_1(a),\dots,P_n(a),\overbrace{0,\dots,0}^{n-k},\overbrace{1\dots,1}^k) \]
Therefore this is the $k-th$ order function (the detailed proof is in appendix \ref{combining appendix}). 

\endproof

We are particularly interested in the \textbf{middlemost-cumulative}. When $n$ is odd, this is the order-cumulative defined by the median order-function. When $n$ is even, we have two middlemost-cumulatives: the lower $\Psi^{\frac{n}{2}}$ and the upper $\Psi^{\frac{n}{2}+1}$. It can be shown that they are \textbf{welfare maximizers} if experts' utilities are measured using the $L_1$ distance in $\mathcal{C}$ to the peak.

%\import{texgraphs/}{summary_tab}

\section{Can experts all contribute to the outcome?}\label{section diversity}

Let us start with an example. If there are three experts. Two of them believe that $a_1 \in \Lambda$ will occur almost surely and the third one believes that $a_2 \in \Lambda$ will occur almost surely, then any order-cumulative (as well as any Level-SP method that satisfies certainty and plausibility preservation) will output a probability function that selects $a_1$ almost surely or $a_2$ almost surely, never a mix of the two.

This lack of diversity in the output may sometimes be non acceptable. For example, when all experts have equal weights and are equally competent, one may feel that the output where alternative $a_1$ has probability $2/3$ to be chosen and alternative $a_2$ has probability $1/3$ of being chosen is a better aggregation. 

More generally, one may wish that society's probability support contains all experts' probability supports. This would means that all opinions are represented with some probability in the aggregation. We are going to formulate a weak diversity axiom, defined for single-minded experts, for which we can characterize a unique Level-SP method. Then, we will show an impossibility result if one desires a stronger form of diversity. 

\subsection{Characterizing the weighted-proportional-cumulative.}

Let us start by defining formally what it means for an expert to be single-minded.

\begin{definition}[Single-minded]
A dirac mass $\delta_a$ is the probability law where alternative $a$ is selected almost surely. An expert is single-minded if their input is a dirac mass.
\end{definition}

The next axiom provides a weak notion of diversity. Experts will have some given weights $(w_i)_{i\in N}$, which is the case in many practical applications. When they are all single-minded they each contribute for exactly their weight.

\begin{axiom}[Weighted proportionality]
If experts are single-minded ($p_i = \delta_{a_i}$ for all $i\in N$),
the aggregation must coincide with the weighted average:
\[\forall (a_1,\dots,a_n)\in \Lambda^N \ \  \psi(\delta_{a_1},\dots,\delta_{a_n}) = \sum_{i\in N} w_i \delta_{a_i};\]
where $w_i \geq 0$ is the weight of expert $i\in N$ with $\sum_{j\in N} w_j=1$.
\end{axiom}

The weighted average $\psi(p_1,\dots,p_n):= \sum_i w_i p_i$ satisfies weighted proportionality but it is not Level-SP. The next theorem shows that there is exactly one PAF satisfying Level-SP and weighted proportionality. This function happens to be certainty preserving and can be uniquely described by a single one-SP voting rule $\mu_{\textbf{w}} : [0,1]^n \rightarrow [0,1]$ defined as follows:
\[\forall \textbf{r}=(r_1,...,r_n), \mu_{\textbf{w}}(\textbf{r}) :=\sup\left\{y \ \middle|\  \sum_{i : r_i \geq y}w_i \geq y \right\}.\]

Before we state the main theorem, let us give an equivalent ``median'' formulation of $\mu_{\textbf{w}}$ when the weights are rational numbers.

\begin{proposition}[Rational weights]\label{Prop_rational_weights}
If all the weights are rationals $w_i = s_i/d \in \mathds{Q}$ then:
\[\forall \textbf{r}=(r_1,...,r_n), \mu_{\textbf{w}}(\textbf{r}) :=\operatorname{median}(\overbrace{r_1,\dots,r_1}^{s_1},\dots,\overbrace{r_n,\dots,r_n}^{s_n},0,1/d,\dots,1-1/d,1) \]
\end{proposition}

\proof{Proof}
The proof can be found in the appendix \ref{appendix rational median}. 

 \endproof

\begin{theorem}[Weighted-Proportional-Cumulative]\label{theorem-proportional-cumulative}
There exists a unique PAF $\psi:\mathcal{P}^n\rightarrow \mathcal{P}$ that verifies Level-SP and weighted-proportionality. It is the unique certainty preserving PAF associated with the one-SP rule $\mu_{\textbf{w}}$: 
\[\forall a\in \Lambda,(P_1,\dots,P_n) \in \mathcal{C}; \Psi(P_1,\dots,P_n)(a) := \mu_{\textbf{w}}(P_1(a),\dots,P_n(a)).\]
%Let us name $\psi$ the weighted-proportional-cumulative.
\end{theorem}

\proof{Proof}
The proof can be found in the appendix \ref{appendix main weighted}.  
\endproof

%{\color{blue}
%\begin{remark}
 %   It follows from the previous that the weights weighted proportional cumulative satisfies the weight properties from section \ref{section weighted experts}. No matter the weights chosen, the grading curve $\gamma : x \rightarrow x$ (the identity) provides a satisfying WAPF.
%\end{remark}
%}

We named the PAF given by the previous theorem the weighted-proportional-cumulative. One of the interesting aspects of the weighted proportional-cumulative is how it beautifully aggregates dominated inputs. The next proposition shows that, in that case, all the experts will contribute in the aggregation, proportionally to their weights, on the segment that best describes their role in the group. In the example of Figures \ref{prop weighted cumulative} and \ref{prop weighted cumulative bis}, there are 3 experts, with weights 0.3, 0.5 and 0.2 respectively. The input is a dominated profile where $P_1$ (in blue) is dominating  $P_2$ (in red) which is dominating $P_3$ (in brown). \footnote{We can see it clearly on the left hand side (Figure \ref{prop weighted cumulative}), with the CDFs but not so easily with the probability density functions (PDFs).} The green line is the weighted cumulative that is better illustrated on the right hand side (Figure \ref{prop weighted cumulative bis}). The weighted-proportional-cumulative first follow the PDF of the first expert for 0.3 of the total mass, then it follows the second expert PDF for 0.5 of the total mass, finally it follows the last expert for the remaining 0.2 of the total mass.  This rule of computation can be generalized as the following proposition shows.
 
 \begin{figure}
    \centering
\begin{minipage}{.48\textwidth}
  \centering
\begin{tikzpicture}[scale=0.9]
\begin{axis}[
    legend pos=south east,
    axis lines = left,
    xlabel = $x$,
    ylabel = $P_i(x)$,
]

\addplot [
    domain=0:5, 
    samples=100, 
    color=blue,
    ]
    {1-exp(-2*x)};
\addlegendentry{$P_1=1-exp(-2x)$}

\addplot [
    domain=0:5, 
    samples=100, 
    color=red,
]
{1-exp(-x)};
\addlegendentry{$P_2=1-exp(-x)$}

\addplot [
    domain=0:5, 
    samples=100, 
    color=brown,
    ]
    {1-exp(-0.5*x)};
\addlegendentry{$P_3=1-exp(-0.5x)$}

\addplot[
    domain=0:0.18, 
    samples=100, 
    style={ultra thick},
    color=green]{1-exp(-2*x)};
    ]

\addplot[
    domain=0.18:0.35,
    samples=10,
    style={ultra thick},
    color=green]{0.3};
]

\addplot[
    domain=0.35:1.6094,
    samples=10,
    style={ultra thick},
    color=green]{1-exp(-x)};
]

\addplot[
    domain=1.6094:3.2188,
    samples=10,
    style={ultra thick},
    color=green]{0.8};
]

\addplot[
    domain=3.2188:5,
    samples=30,
    style={ultra thick},
    color=green]{1-exp(-0.5*x)};
\addlegendentry{$\Psi(\textbf{P})$}
]

\draw[dashed] (0,0.3)--(5,0.3);
\draw[dashed] (0,0.8)--(5,0.8);

\end{axis}
\end{tikzpicture}
\captionof{figure}{Weighted proportional cumulative (in green) with weights $w_1=0.3$, $w_2=0.5$ and $w_3=0.2$ (in cumulative space).}
\label{prop weighted cumulative}
\end{minipage}
\hfill
\begin{minipage}{.48\textwidth}
\centering
\begin{tikzpicture}[scale=0.9]
\begin{axis}[
    legend pos=north east,
    axis lines = left,
    xlabel = $x$,
    ylabel = $p_i(x)$,
]

\addplot [
    domain=0:5, 
    samples=100, 
    color=blue,
    ]
    {2*exp(-2*x)};
\addlegendentry{$p_1=2exp(-2x)$}

%Here the blue parabloa is defined

\addplot [
    domain=0:5, 
    samples=100, 
    color=red,
    ]
    {exp(-x)};
\addlegendentry{$p_2=exp(-x)$}

%Below the red parabola is defined
\addplot [
    domain=0:5, 
    samples=100, 
    color=brown,
]
{0.5*exp(-0.5*x)};
\addlegendentry{$p_3=0.5exp(-0.5x)$}

\addplot[
    domain=0:0.18, 
    samples=30, 
    style={ultra thick},
    color=green]{2*exp(-2*x)};
    ]

\addplot[
    domain=0.18:0.3566,
    samples=10,
    style={ultra thick},
    color=green]{0};
]

\addplot[
    domain=0.3566:1.6094,
    samples=10,
    style={ultra thick},
    color=green]{exp(-x)};
]

\addplot[
    domain=1.6094:3.2188,
    samples=10,
    style={ultra thick},
    color=green]{0};
]

\addplot[
    domain=3.2188:5,
    samples=30,
    style={ultra thick},
    color=green]{0.5*exp(-0.5*x)};
\addlegendentry{$\psi(\textbf{p})$}
]

\draw[green, thick](0.18,1.4)--(0.18,0);
\draw[green, thick](0.3566,0.7)--(0.3566,0);
\draw[green, thick](1.6094,0.2)--(1.6094,0);
\draw[green, thick](3.2188,0.1)--(3.2188,0);

\end{axis}
\end{tikzpicture}
\captionof{figure}{Weighted proportional cumulative (in green) with weights $w_1=0.3$, $w_2=0.5$ and $w_3=0.2$ (in probability space))
}
\label{prop weighted cumulative bis}
\end{minipage}
\end{figure}

\begin{proposition}
[Proportional-cumulative for dominated profiles]
When $\Lambda$ is an in\-ter\-val, if all experts cumulative inputs $P_i$ are continuous and verify $\forall i\in N$, $P_i \geq P_{i+1}$, then the weighted-proportional Level-SP mechanism $\psi: \mathcal{P}^N \rightarrow \mathcal{P} $ of weight \textbf{w} can be computed for this input profile as follows:
\[\psi(\textbf{p})(a) = \left\{ 
\begin{array}{lll}
p_i(a) & \mbox{if} & \sum_{k\leq i-1} w_k \leq P_i(a) < \sum_{k\leq i} w_k \\
0 & \mbox{else} &
\end{array}
 \right.\]
\end{proposition}\label{proposition dominated prop}

\proof{Proof}
The proof can be found in the appendix \ref{appendix rational median}.  
\endproof

\subsection{Characterizing the proportional-cumulative.}

When all experts have the same weight ($w_i=\frac{1}{n}$ for all $i\in N$), we obtain the proportional cumulative 

\begin{definition}
[proportional-cumulative] When all expert have the same weight, the proportional-cumulative $\Psi: \mathcal{C}^N \rightarrow \mathcal{C}$ is the aggregation method defined as follows:
\[\forall \textbf{P}=(P_1,...,P_n), \forall a \in \Lambda, \Psi(\textbf{P})(a) = \operatorname{median}\left(P_1(a),\dots,P_n(a),\frac{1}{n},\frac{2}{n},\dots,\frac{n-1}{n}\right) \]
\end{definition}

Hence the one-SP rule associated with the proportional cumulative is $$g(r_1,...,r_n)=\operatorname{median}\left(r_1,\dots,r_n,\frac{1}{n},\frac{2}{n},\dots,\frac{n-1}{n}\right).$$

This function is called the uniform median in \cite{ICML2016} and the linear median in \cite{JenningsPhD} (see \cite{MathProgPaper} for more details).

\begin{axiom}[Proportionality]
If experts are single-minded (e.g. $p_i = \delta_{a_i}$ for all $i\in N$), the aggregation must coincide with the average:
\[\forall (a_1,\dots,a_n)\in \Lambda^N \ \  \psi(\delta_{a_1},\dots,\delta_{a_n}) = \sum_{i\in N} \frac{1}{n} \delta_{a_i}\]
\end{axiom}

If $\Lambda$ is finite, this corresponds to the proportionality axiom in \cite{EC2019} in the budgeting problem (that can be interpreted as a probability aggregation problem). We therefore also name it proportionality.

\begin{theorem}[Proportionality]
The proportional cumulative is the unique Level-SP PAF satisfying proportionality.
\end{theorem}

This is an immediate consequence of Theorem \ref{theorem-proportional-cumulative} and Proposition \ref{Prop_rational_weights} with $w_i=\frac{1}{n}$ for $i=1,...,n$.

\subsection{Considering a stronger definition for the diversity axiom.}

One may want to have diversity in the aggregation for all inputs and not only dirac inputs (aka single-minded experts). 

\begin{axiom}[Diversity]
A PAF $\psi$ is diverse if for every input probability profile $\textbf{p}=(p_1,...,p_n)$, the support of $\psi(\textbf{p})$ contains the union of the supports of the probabilities $p_i$, $i=1,...,n$. 
\end{axiom}

The mean $\psi(p_1,...,p_n):=\frac{1}{n}\sum_{i=1}^n p_i $ satisfies diversity but is not Level-SP.

The axiom above implies that whenever a Borel event has a positive probability for at least one expert, so does society. In particular, if all experts agree that some event has positive probability, e.g. strong plausibility must be satisfied. Hence, from Theorem \ref{ impossibility strong_plau} we deduce the following result.

\begin{theorem}[Diversity Impossibility]\label{impossibility diversity}
If $\Lambda$ is an interval, no Level-SP unanimous PAF satisfies diversity.
\end{theorem}

\proof{Proof}
Diversity implies strong plausibility. As such we can simply use the strong plausibility impossibility theorem (theorem \ref{ impossibility strong_plau}).
\endproof

\newpage

\section{Comparing aggregation methods of interest.}\label{section comparing}
\subsection{Comparing the proportional and the middlemost cumulatives.}

In our view, the main drawback of the middlemost-cumulative is its lack of diversity. When the various supports $S_i$ of the experts' subjective probabilities $p_i$ are disjoint intervals $I_i$, or more generally, when the $P_i$'s dominate one another (as in Fig \ref{middlemost cumulative}), the outcome of middlemost-cumulative equals the view of the median-expert (see the green curve in Figure \ref{middlemost cumulative} (CDFs) and Figure \ref{middlemost cumulative bis} (PDFs)). 

The proportional-cumulative takes into account all the experts' views in a natural way. As Figure \ref{prop cumulative} shows, the leftmost expert CDF is followed by the proportional-cumulative for one-third of the probability mass. This is a interesting property. Since they are the leftmost expert, the left third of their probability distribution is what best describes how they differs from the other experts. Similarly, the rightmost expert is followed by the proportional-cumulative for the rightmost third of their opinion, which also best represents how their opinion differs from the group. The middle expert is followed for their middle opinion which again is interesting for the same reason. As such, not only do experts contribute for exactly one $n$-th of the final outcome but proportional-cumulative ensures that the most striking aspects of their opinion (compared to the group) are represented.

Given that the order-cumulatives are the unique anonymous Level-SP PAFs that satisfy certainty and plausibility preservations, the proportional-cumulative must violate one of the axioms. As it is anonymous and certainty preserving (because represented by a unique unanimous one-SP rule: the uniform median), it must violate plausibility preservation. This can be seen in Figure \ref{prop cumulative bis}: the density of proportional-cumulative (in green) for the interval $[1,2]$ is zero while the experts' densities for this interval are all strictly positive. We believe that the violation of this axiom is not problematic because it satisfies a "level" version of plausibility.\footnote{A PAF verifies level-plausibility preservation if when a level event has a positive probability for all experts, so does society.} Hence, if we are interested in problems where the decision-maker's final decision is only based on a level-event, only the "level-plausibility preservation axiom" needs to be satisfied, and it is.

\begin{figure}
\centering
\begin{minipage}{.49\textwidth}
\begin{tikzpicture}[scale=0.9]
\begin{axis}[
    legend pos=south east,
    axis lines = left,
    xlabel = $x$,
    ylabel = $P_i(x)$,
]

\addplot [
    domain=0:5, 
    samples=100, 
    color=blue,
    ]
    {1-exp(-2*x)};
\addlegendentry{$P_1=1-exp(-2x)$}

\addplot [
    domain=0:5, 
    samples=100, 
    color=red,
]
{1-exp(-x)};
\addlegendentry{$P_2=1-exp(-x)$}

\addplot [
    domain=0:5, 
    samples=100, 
    color=brown,
    ]
    {1-exp(-0.5*x)};
\addlegendentry{$P_3=1-exp(-0.5x)$}

\addplot[
    domain=0:5, 
    samples=100, 
    style={ultra thick},
    color=green]{1-exp(-x)};
\addlegendentry{$\Psi(\textbf{P})$}

\end{axis}
\end{tikzpicture}
\captionof{figure}{A dominated cumulative profile $(P_1, P_2, P_3)$. Voter $P_1$ is more biased to the left, $P_3$ more biased to the right, $P2$ is the median voter. The middlemost-cumulative (green) coincide with $P_2$.}
\label{middlemost cumulative}
\end{minipage}
\hfill
\begin{minipage}{0.49\textwidth}
\centering
\begin{tikzpicture}[scale=0.9]
\begin{axis}[
    legend pos=south east,
    axis lines = left,
    xlabel = $x$,
    ylabel = $P_i(x)$,
]

\addplot [
    domain=0:5, 
    samples=100, 
    color=blue,
    ]
    {1-exp(-2*x)};
\addlegendentry{$P_1=1-exp(-2x)$}

\addplot [
    domain=0:5, 
    samples=100, 
    color=red,
]
{1-exp(-x)};
\addlegendentry{$P_2=1-exp(-x)$}

\addplot [
    domain=0:5, 
    samples=100, 
    color=brown,
    ]
    {1-exp(-0.5*x)};
\addlegendentry{$P_3=1-exp(-0.5x)$}

\addplot[
    domain=0:0.2027, 
    samples=30, 
    style={ultra thick},
    color=green]{1-exp(-2*x)};
    ]

\addplot[
    domain=0.2027:0.4054,
    samples=10,
    style={ultra thick},
    color=green]{0.3333};
]

\addplot[
    domain=0.4054:1.0986,
    samples=10,
    style={ultra thick},
    color=green]{1-exp(-x)};
]

\addplot[
    domain=1.0986:2.1972,
    samples=10,
    style={ultra thick},
    color=green]{0.6667};
]

\addplot[
    domain=2.1972:5,
    samples=30,
    style={ultra thick},
    color=green]{1-exp(-0.5*x)};
\addlegendentry{$\Psi(\textbf{P})$}
]

\draw[dashed] (0,0.3333)--(5,0.3333);
\draw[dashed] (0,0.6667)--(5,0.6667);

\end{axis}
\end{tikzpicture}
\captionof{figure}{
%The same opinion profile as in Figure \ref{middlemost cumulative}. 
The proportional-cumulative (green) follows at its left the leftist voter $P_1$ for $\frac{1}{3}$ of the mass, at its right the rightist $P_3$ for $\frac{1}{3}$ of the mass, and at its middle the median voter $P_2$ for $\frac{1}{3}$ of the mass.}
\label{prop cumulative}
\end{minipage}
\end{figure}
\begin{figure}
    \centering
\begin{minipage}{0.49\textwidth}
\centering
\begin{tikzpicture}[scale=0.9]
\begin{axis}[
    legend pos=north east,
    axis lines = left,
    xlabel = $x$,
    ylabel = $p_i(x)$,
]

\addplot [
    domain=0:5, 
    samples=100, 
    color=blue,
    ]
    {2*exp(-2*x)};
\addlegendentry{$p_1=2exp(-2*x)$}

%Here the blue parabloa is defined

\addplot [
    domain=0:5, 
    samples=100, 
    color=red,
    ]
    {exp(-x)};
\addlegendentry{$p_2=exp(-x)$}

%Below the red parabola is defined
\addplot [
    domain=0:5, 
    samples=100, 
    color=brown,
]
{0.5*exp(-0.5*x)};
\addlegendentry{$p_3=0.5*exp(-0.5*x)$}

\addplot[
    domain=0:5, 
    samples=100, 
    style={ultra thick},
    color=green]{exp(-x)};
\addlegendentry{$\psi(\textbf{p})$}

\end{axis}
\end{tikzpicture}
\captionof{figure}{Probabilities densities of CDFs in Figure \ref{middlemost cumulative}}
\label{middlemost cumulative bis}
\end{minipage}
\hfill
\begin{minipage}{0.49\textwidth}
\centering
\begin{tikzpicture}[scale=0.9]
\begin{axis}[
    legend pos=north east,
    axis lines = left,
    xlabel = $x$,
    ylabel = $p_i(x)$,
]

\addplot [
    domain=0:5, 
    samples=100, 
    color=blue,
    ]
    {2*exp(-2*x)};
\addlegendentry{$p_1=2exp(-2x)$}

%Here the blue parabloa is defined

\addplot [
    domain=0:5, 
    samples=100, 
    color=red,
    ]
    {exp(-x)};
\addlegendentry{$p_2=exp(-x)$}

%Below the red parabola is defined
\addplot [
    domain=0:5, 
    samples=100, 
    color=brown,
]
{0.5*exp(-0.5*x)};
\addlegendentry{$p_3=0.5exp(-0.5x)$}

\addplot[
    domain=0:0.2027, 
    samples=30, 
    style={ultra thick},
    color=green]{2*exp(-2*x)};
    ]

\addplot[
    domain=0.2027:0.4054,
    samples=10,
    style={ultra thick},
    color=green]{0};
]

\addplot[
    domain=0.4054:1.0986,
    samples=10,
    style={ultra thick},
    color=green]{exp(-x)};
]

\addplot[
    domain=1.0986:2.1972,
    samples=10,
    style={ultra thick},
    color=green]{0};
]

\addplot[
    domain=2.1972:5,
    samples=30,
    style={ultra thick},
    color=green]{0.5*exp(-0.5*x)};
\addlegendentry{$\psi(\textbf{p})$}
]

\draw[green, ultra thick](0.2027,1.32)--(0.2027,0);
\draw[green, ultra thick](0.4054,0.68)--(0.4054,0);
\draw[green, ultra thick](1.0986,0.35)--(1.0986,0);
\draw[green, ultra thick](2.1972,0.185)--(2.1972,0);

\end{axis}
\end{tikzpicture}
\captionof{figure}{Probability densities of CDFs in Figure \ref{prop cumulative}}
\label{prop cumulative bis}
\end{minipage}
\end{figure}

\subsection{Comparing Level-SP methods with the phantom moving mechanisms.}

In the context of Budget aggregation with a finite number of alternatives $\Lambda=\{a_1,...,a_m\}$, \cite{EC2019} proposed a class of anonymous incentive compatible methods (the phantom moving mechanisms). Their notion of strategy\-proofness corresponds, in our context, to incentive compatibility when the voters have single-peaked preferences measured by the $L_1$ distance to the peak in the probability space $\mathcal{P}$, that we will denote by $\|\circ\|^\mathcal{P}_1$. More precisely, if $p=(p_k)_{k=1,...,m}$ and $q=(q_k)_{k=1,...,m}$ are probability distributions over $\Lambda=\{a_1,...,a_m\}$, their distance in the $\mathcal{P}$ space is computed as $d(p,q)=\|p-q\|^\mathcal{P}_1=\sum_{k=1}^m |p_k-q_k | $. 

A voter is single-peaked with respect to $\|\circ\|^\mathcal{P}_1$ if, whenever $p$ is the voter’s peak and $q$ is society’s output then its utility is $-\|p-q\|^\mathcal{P}_1$.  Let’s call a method $L^{\mathcal{P}}_1$-SP if honesty is the optimal strategy when a voters’s utility is single-peaked with respect to $\|\circ\|^\mathcal{P}_1$.

\begin{theorem}[Certainty preserving $L_1$-SP Impossibility]\label{impossibilty l1 centainty}
When $\#\Lambda \geq 4$, the only certainty preserving $\psi$ that are both Level-SP and $L^\mathcal{P}_1$-SP are the dictatorials.
\end{theorem}

\proof{Proof}
The proof can be found in appendix \ref{proof budget dictatorship.} and for additional results see appendix \ref{lambda = 3}. 
\endproof

 Hence, our methods and the one in \cite{EC2019,GKSA2016} are IC in two different environments, but none is strategically robust in both environments, and no method can be IC in both.
 
\begin{remark}
For $\#\Lambda \geq 4$ and without certainty preservation, there exist infinitely many methods that satisfy Level-SP and $L_1^{\mathcal{P}}-SP$. For $\#\Lambda=3$, any Level-SP PAF is $L_1^{\mathcal{P}}-SP$, but not the converse (see appendix \ref{lambda = 3}).
\end{remark}

\section{Characterizing all Level-SP methods with weighted experts.}\label{section weighted experts}

In most practical applications, experts are given weights. In this section we explore what it means to have weighted experts and characterize the Level-SP methods in that case. We will therefore introduce the notion of weighted probability aggregation function (WPAF): $\varphi: \mathds{R}_+^N/\{0\} \times \mathcal{P}^N \rightarrow \mathcal{P}$. Such a function takes as input the weights of the experts $\textbf{w}=(w_1,...,w_i,...,w_n)\in \mathds{R}_+^N/\{0\}$ as well as their probability distributions $\textbf{p}=(p_1,...,p_i,..., ,p_n)\in \mathcal{P}^N $ before outputting a probability distribution, where $w_i$ is the weight of expert $i$ and $p_i$ its probability input.

The next axiom says that if several experts share the same probability distribution then anyone of them can represent the whole by taking over all their weights without changing the outcome.

\begin{axiom}[W-Additivity.]
A WPAF $\varphi: \mathds{R}_+^N/\{0\} \times \mathcal{P}^N \rightarrow \mathcal{P}$ is weight additive if:

\begin{align*}
    \forall S \subseteq N, \forall \textbf{p}, \forall \textbf{w}, \textbf{w'}: \ (\exists i \in S, \forall j \in S, p_i = p_j) \wedge(\sum_{i \in S} w_i = \sum_{i \in S} w'_i) \Rightarrow  \varphi(\textbf{w},\textbf{p})=\varphi(\textbf{w'},\textbf{p})
\end{align*}
\end{axiom}

The next axiom describes that no expert should benefit from having their weight decreased. As before the utility of experts is assumed to be strongly connected to level events.

\begin{axiom}[W-Monotonicity]
    A WPAF $\varphi: \mathds{R}_+^N/\{0\} \times \mathcal{P}^N \rightarrow \mathcal{P}$ is weight monotone if for all experts $i$, for all weights $\textbf{w}$ and $\textbf{w'}$ that only differ for expert $i$ with $w_i < w'_i$:

\[\forall a \in \Lambda, \forall \textbf{p} \in \mathcal{P},
|\varphi(\textbf{w},\textbf{p})(\mathcal{E}(a)) - p_i(\mathcal{E}(a))| \geq |\varphi(\textbf{w'},\textbf{p})(\mathcal{E}(a))-p_i(\mathcal{E}(a))|. \]
\end{axiom}

The next axiom says that the outcome does not depend by which scale the weights are measured. In other words if all weights are multiplied by a same positive constant k, the outcome is unchanged.

\begin{axiom}[W-Proportionality.]
A WPAF $\varphi: \mathds{R}_+^N/\{0\} \times \mathcal{P}^N \rightarrow \mathcal{P}$ is weight proportional if:
\[\forall \textbf{p} \in \mathcal{P}, \forall \textbf{w}, \forall k \in \mathds{R}_{+}^*,
\varphi(\textbf{w},\textbf{p}) = \varphi(k\textbf{w},\textbf{p}). \]
\end{axiom}

Finally our final axiom requires that experts with the same weight should be treated equally.

\begin{axiom}[W-Anonymity]
A WPAF $\varphi: \mathds{R}_+^N/\{0\} \times \mathcal{P}^N \rightarrow \mathcal{P}$ is weight anonymous if:
\[\forall \textbf{w}, i,j, \ \ \  (w_j=w_i) \Rightarrow \forall \textbf{p}, \varphi({\textbf{w}},(\textbf{p}_{-i}(p_j))_{-j}(p_i))= \varphi(\textbf{w},\textbf{p}).\]
\end{axiom}

A WPAF $\varphi$ is Level-SP if no matter the sets of weights $\textbf{w}$ associated to the experts, the resulting PAF $\psi_{\textbf{w}} (\textbf{p}):= \varphi(\textbf{w},\textbf{p})$ is Level-SP. We denote $f_S^\textbf{w}$, $S\subset N$, the phantom functions associated to $\psi_{\textbf{w}}$.

\begin{theorem}\label{theorem weighted experts}
       A WPAF $\varphi: \mathds{R}_+^N/\{0\} \times \mathcal{P}^N \rightarrow \mathcal{P}$ is Level-SP, W-Additive, W-Proportional, W-Monotone and W-Anonymous iff there exists a function $\gamma: \Lambda \times [0,1] \rightarrow [0,1]$ right continuous and weakly increasing in its first argument, and weakly increasing in its second argument with for all $a$, $\lim_{a \rightarrow \sup \Lambda}g(a,1)=1$, such that $f_S^\textbf{w}$ the phantom function associated to $\psi_{\textbf{w}}$ can be computed as:

    \[f_S^\textbf{w}(a) =\gamma\left(a,\frac{\sum_{j \in S} w_j}{\sum_{j\in N}w_j} \right).\]

    And in that case, $\forall \textbf{w} \in \mathds{R}_+^N/\{0\},$ the cumulative of $\psi_{\textbf{w}}(\cdot):=\varphi(\textbf{w},\cdot)$ can be computed as follows:
    \[\forall a\in\Lambda, \forall \textbf{P} \in \mathcal{C}^N \Psi_{\textbf{w}}(\textbf{P})(a) := \sup \left\{ y \middle| \gamma\left(a,\dfrac{\sum_{i: P_i(a) \geq y}w_i}{\sum_{i \in N}w_i}\right) \geq y \right\}.\] 

\end{theorem}

\proof{Proof}
The proof can be found in appendix \ref{appendix weighted experts}.  
\endproof

\begin{definition}[Weight consistent]
    A PAF $\psi$ is said to be weight consistent if there exists a WPAF $\varphi$ satisfying the four W-axioms above and positive weights $\textbf{w}=(w_1,...,w_n)\mathds{R}_+^N/\{0\}$ such that:
    \[\forall \textbf{p}, \forall a \in \Lambda, \psi(\textbf{p})(a) = \varphi(\textbf{w},\textbf{p})(a).\]

    The WPAF $\varphi$ is said to extend $\psi$.

\end{definition}

Hence as a corollary of the previous theorem, we obtain a full characterization of weight consistent PAFs.

\begin{theorem}
A Level-strategyproof PAF $\psi :\mathcal{P}^N \rightarrow \mathcal{P}$ is weight consistent  %verifies that all experts have weights that are "consistent", "define", "proportional" and "additive" 
    iff there are weights $\textbf{w}=(w_1,...,w_n)$ and a (grading) function $\gamma: \Lambda \times [0,1] \rightarrow [0,1]$ weakly increasing in its second argument such that the phantom functions $f_S$ all verify:

    \[f_S(a) =\gamma\left(a,\frac{\sum_{j \in S} w_j}{\sum_{j\in N}w_j} \right).\]

    or equivalently, iff the cumulative of $\psi$ can be computed as follows:
    \[\forall a\in\Lambda, \forall \textbf{P} \in \mathcal{C}^N \Psi(\textbf{P})(a) := \sup \left\{ y \middle| \gamma\left(a,\dfrac{\sum_{i: P_i(a) \geq y}w_i}{\sum_{i \in N}w_i},y\right) \geq y \right\}.\] 

\end{theorem}

\begin{remark}
    It follows that the weighted proportional cumulative is weight consistent. 
    The grading curve $\gamma : x \rightarrow x$ provides a WAPF that extends it, no matter what the weights are.
\end{remark}

\begin{remark}
    It also follows, from the grading curve characterization in \cite{MathProgPaper},  that anonymous Level-SP PAFs are weight consistents (with equal weights for the experts).
\end{remark}

\begin{remark}
The W-axioms make sense in most aggregation problems, and not only the probability aggregation one.
\end{remark}

%{\color{red} Ou est passer le corollaire 1, C est la propriete principale du paragraphe}

\section{Application: Electing and ranking with uncertain voters.}\label{section MJ and MJU}

The objective of this section is to show that the weighted-proportional-cumulative could be combined with a recent evaluation-based voting method (Majority Judgment \cite{BL2011,Balinski-Laraki}, \textbf{MJ}) to construct a strategically robust voting method in situations where voters are uncertain and have doubts about the candidate pool.

We will first recall MJ and its salient properties before describing its extension to uncertainty.

\subsection{Majority Judgment: A grading method for voting.}
 
Let us first provide a simple example to describe how Majority Judgment (MJ) works in an anonymous setting. Let us suppose the designer has fixed a numerical scale $\{0,\dots,9\}$ and, as example, consider a set of five grades ordered from best to worse $A=(9,7,6,5,2)$ corresponding to the grades given by 5 voters to a candidate (or alternative) A. Its majority value is obtained by iterating from the middlemost grade (the median), then down, up, down, etc, which gives us the five dimensional vector $v(A)= (6,5,7,2,9)$. Two ordered sets of grades $A$ and $B$ are compared in lexicographic order by their majority values. For example if $B=(9,8,6,4,1)$ then $v(B)=(6,4,8,1,9)$ and so $v(A)>v(B)$ because $(6,5,7,2,9)\succcurlyeq(6,4,8,1,9)$. 

If he numerical grades are replaced by some monotonic transformation, or by a qualitative set of grades such as $\{$Great, Good, Average, Poor, Terrible$\}$ the ranking is unchanged. It follows that MJ is is an ordinal method.

%Majority judgment is an ordinal method as the ranking remains unchanged if the numerical grades are replaced by some monotonic transformation, or by a qualitative set of grades such as $\{$Great, Good, Average, Poor, Terrible$\}$. 

Formally, suppose there are $n$ voters who assign each candidate a grade in an ordinal scale $\Lambda$. 

The $k$th \textit{order function} $f^k:\Lambda^n \rightarrow \Lambda$ is the social-grading function whose value is the $kth$ highest grade: \[\textbf{r}=(r_1\succeq r_2\succeq \cdots \succeq r_n)\Rightarrow f^k(\textbf{r})=r_k.\] 
A candidate's \textit{majority-grade} $f^{maj} (\textbf{r})$ is the highest grade approved by an absolute majority:
\[f^{maj}(\textbf{r})=\left\{\begin{array}{ll}f^{\frac{n+1}{2}}(\textbf{r})& \mbox{if $n$ is odd,}\\
f^{\frac{n+2}{2}}(\textbf{r})&\mbox{if $n$ is even.}\end{array}\right.\]

Suppose a candidate's merit profile is \[r_1\succeq r_2\succeq\cdots \succeq r_n.\]
Its \textit{majority-value} is an ordered sequence of these grades. The first in the sequence is its majority-grade; the second is the majority-grade when its (first) majority-grade has been dropped (it is its ``second majority-grade''); the third is the majority-grade  when its first two majority-grades have been dropped; and so on.\footnote{Thus, when there is an odd number of voters $n=2t-1$, a candidate's \textit{majority-value} is the sequence that begins at the middle, $r_{t}$, and fans out alternately from the center starting from below:$\overrightarrow{\textbf{r}}=(r_{t},r_{t+1},r_{t-1},r_{t+2},r_{t-2},\ldots ,r_{2t-1},r_1)$. When there is an even number of voters $n=2t-2$, the majority-value  begins at the lower middle and fans out alternatively from the center starting from above $\overrightarrow{\textbf{r}}=(r_t,r_{t-1},r_{t+1},r_{t-2},r_{t+2},\ldots ,r_{2t-2},r_1).$
} The MJ ranking can be extended to a continuum of voters (e.g. a normalized distribution over the set of grades, see \cite{BL2011} chapter 14).

\begin{figure}
\centering
\begin{minipage}{.51\textwidth}
  \centering
  \includegraphics[width=\textwidth]{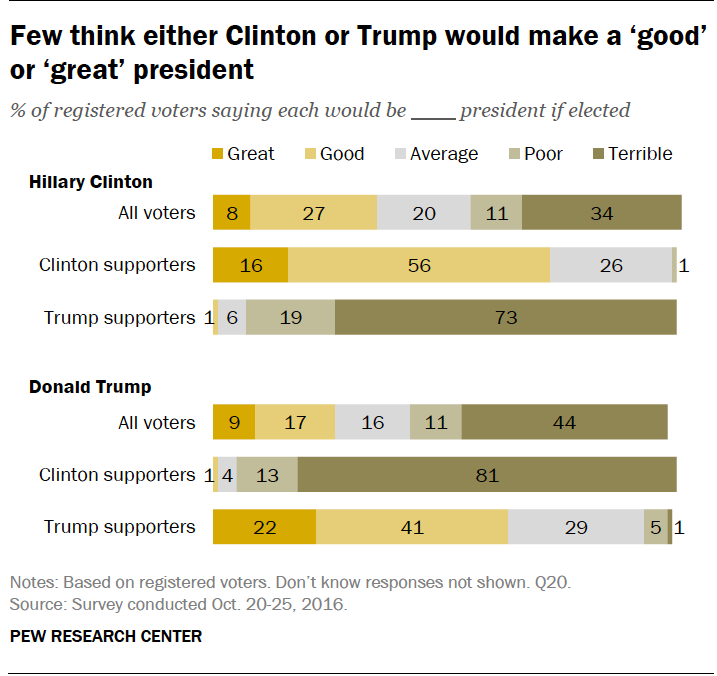}

  \label{hilary}
\end{minipage}%
\begin{minipage}{.1\textwidth}
  
\end{minipage}
\begin{minipage}{.48\textwidth}
  \centering
 \includegraphics[width=\textwidth]{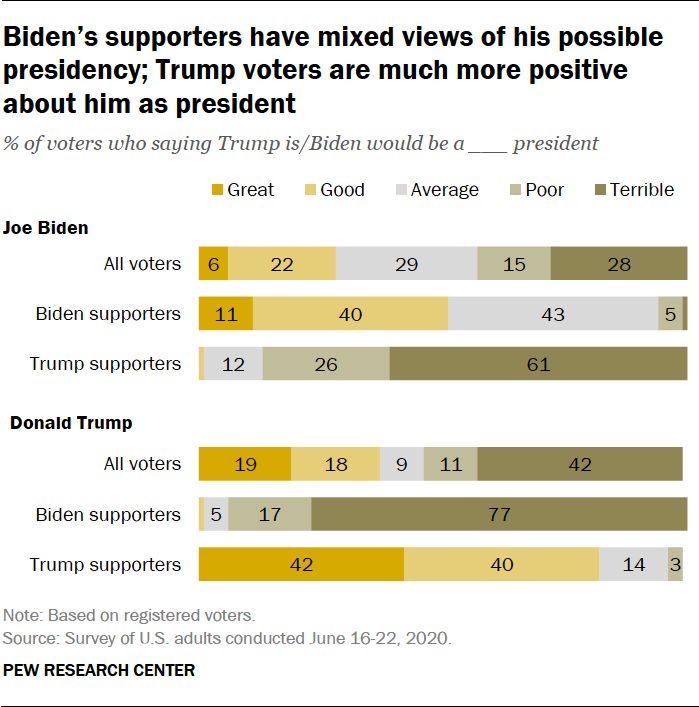}
  \label{biden}
\end{minipage}

\caption{\footnotesize{In majority judgment \cite{Balinski-Laraki,BL2011}, each voter is asked to grade each candidate on a scale $\Lambda$ such as ``Great, Good, Average, Poor, Terrible''. The output for each candidate (the merit profile) can be interpreted as a probability distribution over $\Lambda$.  The winner is the one with the highest median grade (the majority-grade). In case of the same majority-grade, an elaborate iterative rule breaks the tie (see the main text for a formal definition). In the surveys above, Donald Trump's majority grade is ``Poor'' in both elections and so with majority judgment they lose against both candidates, Clinton and Biden, since their majority grade is ``Average''. The two figures are extracted from two Pew Research Center survey reports about the 2016 and the 2020 US elections \cite{Pew1,Pew2}.}}
\end{figure}

\medskip

Let us now extend MJ to settings where voters have weights. It can be formally justified using the W-axioms.

\noindent\textbf{A formal description of MJ when voters have weights}:
Let us imagine $N$ voters who should rank $M$ candidates or alternatives $\mathcal{A}=\{A_1, A_2, A_3,\dots,A_M\}$. Majority Judgment works as follows:
\begin{itemize}
\item \textit{Fixed by the designer:} a scale of grades $\Lambda$ (such as $\{$Great, Good, Average, Poor, Terrible$\}$ or 
$[0,100] $) and a normalized positive vector $\mathbf{w}=(w_1$,\dots,$w_N)$ where $w_i$ is the weight of voter $i$.
\item \textit{Input from voters (the ballots):} each voter $i\in N$ is asked to submit for each candidate $A\in \mathcal{A}$ the grade in $\Lambda$ they associate to that candidate.
\item \textit{Output for voters:} each candidate $A$ is given a normalized distribution $p^A$ over the set of grades $\Lambda$ (that may be seen as a probability distribution over $\Lambda$) where the probability of each alternative is the sum of the weights of the voters that voted for it.
\item \textit{Ranking of the alternatives:} For any two alternatives $A$ and $B$, we rank them by using algorithm \ref{algo:main}. Basically the majority grade given to each candidate $A$ is the value of $(P^A)^{-1}(0,5)$. When two candidates have the same majority grade, we look at what happens in the closest point to $0,5$ where they disagree.

\end{itemize}

\begin{algorithm}
\begin{algorithmic}
\If{$p^A = p^B$}
\State $p^A =_{MJ} p^B$ 
\EndIf
 
\State $a \gets \sup(\{x \leq 0,5 | (P^A)^{-1}(x) \neq (P^B)^{-1}(x)\} \cup \{0\})$\\
\State $b \gets \min(\{x \geq 0,5 | (P^A)^{-1}(x) \neq (P^B)^{-1}(x)\}\cup \{1\})$ \\
\If{$b-0,5 < 0,5 -a$ or $b=0,5$}
\State $c \gets b$
\Else
\State $c \gets a$
\EndIf
\If{$c=b$}
\If{$(P^A)^{-1}(c) < (P^B)^{-1}(c)$}
\State $p^A <_{MJ} p^B$
\Else
\State $p^A >_{MJ} p^B$
\EndIf
\Else
\If{$\exists \epsilon>0$ s.t. $\forall x \in [a-\epsilon,a[,(P^A)^{-1}(x) < (P^B)^{-1}(x)$}
\State $p^A <_{MJ} p^B$
\Else
\State $p^A >_{MJ} p^B$
\EndIf
\EndIf
\end{algorithmic}
\caption{Ranking A and B in MJ and MJU}
\label{algo:main}
\end{algorithm} 

\begin{remark}
    The ranking is well defined since for each candidate $A$, we have $\#Im(P^A) \leq N$.
\end{remark}

%Let us now recall some of the interesting properties of Majority Judgment.

\noindent\textbf{The salient properties of MJ \cite{BL2011,Balinski-Laraki} }:

\begin{enumerate}
    \item \label{MJ a} MJ outputs a transitive order (and so it avoids the Condorcet paradox);
    \item \label{MJ b} MJ satisfies IIA (Independence of Irrelevant Alternatives): adding or dropping a candidate does not change the ranking between the others. Hence, MJ is not subject to the Arrow paradox \cite{Arrow}.
    %the Arrow paradox \cite{Arrow} is avoided by MJ;
    \item \label{MJ c} Candidates are treated equally;
    \item \label{MJ d} When voters have the same weights, their ballots are treated equally;
    \item \label{MJ e} MJ best resists strategic manipulations among the methods satisfying the above properties. This is shown in the exam below and empirically (see \cite{BL2011}, chapter 19).
\end{enumerate}

%{\color{red} See proof? illustration ou preuve de \ref{MJ e}. X puis A puis B?}

For example, if a voter gives a candidate X a grade higher than X's  majority grade, they cannot increase the  majority-grade of X, and symmetrically if they gave a grade lower than X's majority grade they cannot lower X's majority grade. Moreover, if a voter prefers a candidate A to a candidate B, and B is ahead of A by MJ, then they cannot decrease B's  majority grade and symmetrically cannot increase A's majority grade (e.g. MJ method is partially strategyproof in ranking). Also, while no method satisfying \ref{MJ a}, \ref{MJ b} and \ref{MJ c} is strategyproof in ranking, MJ is the unique method satisfying \ref{MJ a}, \ref{MJ b} and \ref{MJ c} which is strategyproof in ranking on a subdomain where it coincides with Condorcet's method (see \cite{Balinski-Laraki}).

%Most voting methods (plurality, approval voting, Borda, single transferable vote, range voting, MJ, etc) elect the Condorcet winner whenever it exists under several Nash equilibrium refinements. With MJ, additionally, the (Condorcet) winner's majority grade is its honest majority grade and most of the grades in the refined Nash equilibria are honest (see \cite{BL2011}, chapter 20) {\color{red} Phrase dure a lire}. This is a nice property because the objective of an election is not only to elect a candidate but measure the true evaluation of the candidate in the society and thus it is important that the method output the honest majoritarian judgment.        

\subsection{Majority Judgment with uncertainty (MJU).}

To elect one candidate/alternative or rank several ones \cite{Arrow}, existing voting methods (plurality, Borda, Condorcet, approval voting, etc) implicitly assume that individual voters are certain about their opinions or views. In practice, voters are mostly uncertain. In recruitment committees, members are often hesitating between a good safe candidate vs a risky one. On the day of the Brexit vote, no voter knew with certainty what the final deal between the UK and the EU would be, nor did they know the long-term consequences of such a deal. Similarly, when a reviewer is judging a paper for a conference, they are often uncertain about the quality of some of the papers. 

%To capture those uncertainties, one can use an extension of majority judgment.

The philosophy behind MJ is to allow voters to better express themselves (compared to plurality or approval voting for instance) by submitting a grade on a scale $\Lambda$ for each candidate. MJU goes further in this philosophy by allowing voters to express their uncertainties about the candidates on the same scale.

\newpage

\noindent\textbf{A formal description of MJU}:
\begin{itemize}
\item \textit{Fixed by the designer:} a \textbf{finite} scale of grades $\Lambda$ (such as $\{$Great, Good, Average, Poor, Terrible$\}$) and a normalized positive vector $\mathbf{w}=(w_1$,\dots,$w_N)$ where $w_i$ is the weight of voter $i$.
\item \textit{Input from voters (the ballots):} each voter $i\in N$ is asked to submit for each candidate $A\in \mathcal{A}$ a probability distribution $p_i^A \in \mathcal{P}=\Delta(\Lambda)$. For example, the voter may think that $A_k$ will be Good for sure, and that $A_l$ will be Great with probability $\frac{2}{3}$ and Terrible with probability $\frac{1}{3}$.
\item \textit{Output for candidates:} each candidate $A$ is given an aggregate probability distribution $p^A$ \textbf{computed using the weighted-proportional-cumulative} that is  $P^A(a)=\mu_{\mathbf{w}}(P^A_1(a),\dots,P^A_N(a))$.
\item \textit{Ranking of the alternatives:} classical \textbf{majority judgment is applied} to rank the distributions $p^{A_1}$, $p^{A_2}$,\dots, $p^{A_M}$ (see Algorithm \ref{algo:main}) and consequently the candidates/alternatives in $\mathcal{A}$. 
\end{itemize}

%The table represents the votes for the MUJ.

\begin{table}
    \centering
    \begin{tabular}{|c|c|c|c|c|c|}
    \hline
          & Great & Good & Average & Poor & Terrible  \\
    \hline
         Judge 1's probability distribution for candidate A & 41 & 46 & 3 & 5 & 5 \\
         Judge 2's probability distribution for candidate A & 20 & 20 & 20 & 20 & 20 \\
         Judge 3's probability distribution for candidate A & 54 & 21 & 2 & 4 & 19 \\
         Proportional cumulative for candidate A & 41 & 34 & 2 & 4 & 19 \\
\hline 
         Judge 1's probability distribution for candidate B & 8 & 18 & 34 & 19 & 21 \\
         Judge 2's probability distribution for candidate B & 32 & 9 & 20 & 28 & 11 \\
         Judge 3's probability distribution for candidate B & 50 & 0 & 0 & 0 & 50 \\ 
         Proportional cumulative for candidate B  & 32 & 9 & 19 & 19 & 21 \\
\hline
         Judge 1's probability distribution for candidate C & 30 & 0 & 50 & 16 & 4 \\
         Judge 2's probability distribution for candidate C & 13 & 33 & 28 & 2 & 34 \\
         Judge 3's probability distribution for candidate C & 0 & 35 & 42 & 5 & 18 \\
         Proportional cumulative for candidate C & 13 & 28 & 39 & 2 & 18 \\
\hline
    \end{tabular}
    \caption{An example of MJU with three equal weights voters, three candidates, and  where candidate A wins.}
    \label{simpleMJU}
\end{table}

Let us give an intuition of MJU with the simple anonymous example provided in table \ref{simpleMJU}. The first line represents the probability distribution of judge 1 about candidate A, hence judge 1 believes that with 41\% chance A will be great candidate, with 46\% chance A will be a good candidate and so one. The fourth line is the final distribution of merits of candidate A computed by applying the proportional-cumulative to aggregate judges 1, 2 and 3 distributions for A. To rank A, B and C, observe that candidate A gets the majority grade ``good'' and Candidate B and C both get majority grade ``average''. As such, A ranks better than B and C. It remains therefore to break the tie between B and C. According to the proportional-cumulative, there is a 40\% chance that B is strictly less than ``average'' compared to 20\% for C. They both have 41\% chance of being strictly better than ``average''. As such C is ranked above B. The final ranking is therefore $A > C > B$.

\medskip

\noindent\textbf{The salient properties of MJU}

\begin{proposition}
    MJU induces an order (hence, the method avoids Condorcet's paradox).
\end{proposition}

\proof{Proof}
Let us show that the ranking method verifies the definition of a partial order.
    \begin{itemize}
        \item Reflexive: We have $p^A \leq_{MJ} p^A$.
        \item Transitive: The condition $p^A <_{MJ} p^B$ can be described as the existence of an interval $[a,b[$ centered in $0,5$ on which $(P^A)^{-1} \leq (P^B)^{-1}$ with at least one point $c$ such that $P^A(c) < P^B(c)$. Similarly the condition $p^A =_{MJ} p^B$ can be considered to mean that $P_A = P_B$. Therefore if $p^A \leq_{MJ} p^B$ and $p^B \leq_{MJ} p^C$ then by choosing the smallest of the two intervals defined by these inequalities we find an interval that shows that $p^A \leq_{MJ} p^C$,
        \item Asymmetrical. Suppose that $p^A \leq_{MJ} p^B$ and $p^B \leq_{MJ} p^A$. Then by the conditions we have that for all intervals centered on $0,5$ we have $(P^A)^{-1}=(P^B)^{-1}$. As such $p^A =_{MJ} p^B$.
    \end{itemize}

    It remains to show that our ordering is total when the number of grades is finite. Since there is a finite number of grades there is a finite number of points were either $(P^A)^-1$ or $(P^B)^-1$ changes value. The existence of a $c$ and possibly a $\epsilon$ that defines the ranking is therefore ensured. 
    
    \endproof

\begin{remark}
    Unlike in MJ, we only consider finite scales. With an infinite scale, we only get a partial order over candidates as the $\epsilon>0$ sometimes needed in algorithm \ref{algo:main} to break ties does not always exist.
\end{remark}

\begin{proposition}{[IIA]}
    In MJU, adding or dropping a candidate does not change the ranking between the others candidates  (the method avoids Arrow's paradox)
\end{proposition}

\proof{Proof}
    The ranking between two candidates $A$ and $B$ by MJU depends only on their distributions $p^{A}$ and $p^{B}$. As for the distribution of a candidate, it only depends on the votes associated to that candidate. Adding a new candidate doesn't affect any of these. 
    
    \endproof

\begin{proposition}{[Impartiality]}
    In MJU, all candidates are treated equally.
\end{proposition}

\proof{Proof}
    We apply the same PAF to all candidates. As such, for the same input they get the same aggregated probability distribution.It follows that the process to determine the aggregated probability distribution is impartial. It remains to show that the ranking method of probability distributions is also impartial. The order used by the ranking method only depends on the induced distributions that it is comparing. It is therefore impartial. 
    
    \endproof

An important reason of using the proportional cumulative (a Level-SP method) to aggregate the probabilities received by a candidate is to reduce the incentive of a voter to lie about the probability distribution they give to a candidate. 
MJ is an iteration of the majority grade operator, and the majority grade is defined via level events. It follows that to resist to strategic manipulations, one should use a Level-SP method. More formally, we can prove the following proposition, which extends a similar result that exists for MJ.

\begin{proposition}{
    [Partial strategyproofness in ranking]} 
   If the proportional cumulative (or any Level-SP method) is used to aggregate voters probability's inputs at the candidate level and if candidate $A$'s society majority grades is strictly above candidate $B$'s society majority grade, and if voter $i$'s input probability distribution for $B$ (first order stochastically) dominates $i$'s input probability distribution for $A$ then:
   \begin{itemize}
       \item if $i$ can increase society's majority-grade of B, they cannot decrease that of A;
       \item if $i$ can decrease society's majority-grade of A, they cannot increase that of B.
   \end{itemize}
\end{proposition}

\proof{Proof}
For a candidate $X=A,B$, let $m(X)$ denote society's majority grade of $X$, and $m_i(X)$ be the majority grade of voter $i$ for $X$. 

Suppose $m_i(B)>m(B)$. If $\alpha$ is the grade just above $m(B)$ in the scale $\Lambda$ then $P_i(\alpha)\geq P_i(m_i(B)) \geq 1/2$, and if $P(\alpha)$ denote society's cumulative at $\alpha$, then since $\alpha > m(B)$ and $m(B)$ is society's majority grade for B, we must have  $P(\alpha)<1/2$. Consequently, $P_i(\alpha)>P(\alpha)$ and because we are using a Level-SP method, voter $i$ cannot increase the cumulative at $\alpha$ of candidate $B$ and consequently they cannot increase society's majority grade of B. Hence, if voter $i$ can increase $B$'s majority grade, we must have that $m_i(B)\leq m(B)$, but since our assumptions imply that $m(A)>m(B)$ and $m_i(B)\geq m_i(A)$ we obtain that $m_i(A) < m(A)$ and thus, by an argument symmetrical to the one above, we deduce that voter $i$ cannot decrease society's majority grade of $A$.\footnote{The proof only uses that $m_i(B) \geq m_i(A) $. This condition is satisfied whenever $i$'s distribution for $B$ dominates its input for $A$, but there are other cases in which $i$ prefers $B$ to $A$ for which this condition is still satisfied.} 

\endproof

Why is the previous theorem so interesting and important? If society's majority grade of $A$ is higher than that of $B$, MJ ranks $A$ above $B$. Now, if voter $i$ prefers $B$ to $A$ (for example if the probability distribution for $B$ dominates the one for $A$), they may be tempted to vote strategically in order to inverse society's ranking. If they cannot increase $B$'s majority grade nor decrease $A$'s majority grade then $i$ cannot manipulate the ranking between $A$ and $B$ at all. This is the case whenever $m_i(B)>m(b)$ and $m_i(A)<m(A)$ which is the most likely case in practice (because if $i$ is a partisan of $B$ and an opponent of $A$, their majority grade for $B$ (resp. $A$) will typically be higher (resp. lower) than society's majority grade of $B$ (resp. $A$), see \cite{Balinski-Laraki} for a practical instance). Otherwise (from the proof just above), if $m_i(B)\leq m(b)$ then our assumptions imply that $m_i(A)<m(A)$ and so voter $i$ cannot decrease $A$'s society majority grade and if  $m_i(A) \geq m(A)$ then $m_i(B)>m(b)$ and so $i$ cannot increase $B$'s society majority grade. In sum, either a voter cannot manipulate or they can only partially manipulate (e.g. only in one direction).

%{\color{red} I do not agree about it being the most likely case.}

The reason why we should use the proportional cumulative and not another Level-SP method is because, among the Level-SP rules, it is the only one that guarantees that the MJU ranking coincides with the MJ ranking when all the voters are single-minded.

\begin{proposition}{[Extends MJ]}
    If all voters are single-minded (for each candidate their input is a Dirac in one of the grades), then the distributions and the rankings of the candidates obtained by MJU are the same as the one obtained by MJ when the inputs are single-minded (Dirac distributions). 
\end{proposition}

This proposition is key because the objective of all this exercise is to extend MJ to uncertainty and so to coincide with MJ when voters are 100\% certain about the qualities of the candidates.

\proof{Proof}
    The proof is immediate with the fact that the aggregated probability distribution in the single-minded case is the same in MJ and MJU because MJU aggregate the voters probabilities at the candidate level using the weighted proportional cumulative which satisfies the weighted-proportionality axiom. 
    
\endproof

To extend MJ, one could have used any other PAF that satisfies weighted-proportionality (such as the weight average $\psi (p_1,...,p_n)=\sum_i w_i p_i$). However, the weighted-proportinal-cumulative is the only weighted-proportional and Level-SP PAF. Consequently any other weighted-proportional PAF would not be partially-strategy proof in ranking, as can be proved. 
%{\color{red}C vraiment trivial a ce point?}

%a rule, different from the weighted proportional cumulative, will not be Level-SP and consequently, the ranking method will not be partially-strategy proof in ranking, as one can easily check.

What about the extension of other voting methods to uncertainty? Arrowian voting methods \cite{Arrow} based on ranking-orderings --e.g. voters are asked their full preference ordering over all the candidates (Borda, STV)-- cannot easily be extended to uncertainty, at least in a practical way. A possible extension is to submit a probability over the set of permutations over $\mathcal{A}$, e.g. with probability $\frac{1}{2}$ the ranking is $A>B>C>D$ and the probability $\frac{1}{2}$ it is $C>B>A>D$? Theoretically, this can be done and several known methods can be consistently extended to uncertainty. In practice, it seems a very difficult task for voters as the interpretation is not clear. How does a voter decide on his vote when uncertain about a few candidates?. Moreover, the dimension of the space of probabilities is very high ($m!-1$ with $m$ candidates). This can lead to mistakes and hesitations among voters. Though future experimentation is necessary to determine how true this affirmation is. Approval voting, which asks each voter to submit a subset of the candidates they approve, can be extended to uncertainty in two ways. A first method would be for voters to submit a probability distribution over subsets of the candidates: e.g. with probability $\frac{1}{2}$ the voter approves $\{A,B,C\}$ and with probability $\frac{1}{2}$ they approve $\{E,F,G\}$. This distribution looks curious and the exercise would be complex as the dimension of the probability space is exponential ($2^m-1$, with $m$ candidates). The second (simpler) method is to submit for each candidate and independently from the others, its approval probability (e.g. $i$ approves each candidate $A\in \mathcal{A}$ with probability $p^i_A$, and so they submit only $m$ numbers in $[0,1]$). The interpretation is that Approve/Disapprove has some absolute meaning (e.g. Good/Bad, Innocent/Guilty) and the probability is subjective opinion of the voter about the status of the candidate/alternative (e.g. the voter believes the candidate will be a good president with probability $\frac{1}{2}$ and bad president with probability $\frac{1}{2}$). In that case, a consistent extension of AV to uncertainty is to use the MJU rule on a $\Lambda$ with only two grades. We can call this method AVU (Approval Voting under Uncertainty). 

%But with only two grades in $\Lambda$, proportional-cumulative coincides with uniform median $\mu$.
%But as observed in section 4, with only two grades in $\Lambda$, proportional-cumulative coincides with the IMM and which coincides with the Single-Dimension-SP median rule with uniformly distributed phantoms. 
%In the AVU, each candidate will receive a $p^A$ probability of Approval 

%To end this discussion, 

Other methods can be extended. For example, consider binary decision problems such as a referendum (e.g. the Brexit), and extend majority rule  ---with the Condorcet Jury context in mind--- by allowing voters to express their uncertainties (see \cite{TBK0} for a similar approach). More precisely, each voter is asked to report its prior probability that some reform (e.g. the Brexit) is the right decision compared to the status quo. This is represented by a number $p_i\in [0,1]$ for every voter $i=1,\dots,n$. Then aggregate those numbers using proportional-cumulative which coincides with the uniform median rule: $$p=\mu(p_1,\dots,p_n)=\operatorname{median}\left(p_1,\dots,p_n,0, \frac{1}{n},\frac{2}{n},\dots,\frac{n-1}{n},1\right)= \sup_{q \in [0,1]} \left\{ \frac{\#\{i\in N: p_i \geq q\}}{n} \geq q \right\}. $$
Finally, if $p>1/2=\alpha$ then the reform is implemented, otherwise, it is not and we keep the status quo. Observe that when all voters are single-minded and vote extreme values ($p_i \in \{0, 1\}$ for all $i\in N$), then $p=\# \{i\in N: p_i=1\}/n$, and so the procedure defines an extension of majority rule.
%(that we use a method that satisfies the proportionality axiom is, of course, crucial to have its extension). 
But, as the consequences of making a bad decision can be high compared to the status quo (e.g. the Brexit), one may argue that the optimal $\alpha$ must be higher than $\frac{1}{2}$, perhaps $\frac{2}{3}$? In theory, the optimal $\alpha$ can be calculated by the maximization of some vNM expected utility of society. That is, we normalize the society vNM utility to 1 if the reform is a good decision and to 0 if the reform is a bad decision, then the optimal $\alpha$ coincide with the vNM utility of the status quo. In practice, the optimal $\alpha$ can be fixed by some committee (e.g. the house of commons) or it can be determined democratically as the median of society individual risk aversions $\alpha=\operatorname{median}(\alpha_1,\dots,\alpha_n)$.

%Defined as such, MJU inherits some of resistance to manipulation properties of MJ.

%This is the same type of ideas as with MJ were cannot get a grade that is closer to our favorate peak.

\section{Conclusion.}\label{conclusion}

This paper studies the probability aggregation problem when the set of outcomes $\Lambda$ is an ordered set. It defines level-strategy\-proofness and proves it to imply classical strategy\-proofness for a rich set of single-peaked preferences over the CDF space. Several characterizations are established when level-strategy\-proofness is combined with other axioms and two methods are singled out: middlemost-cumulative and proportional-cumulative. Both are easy to compute and can be extended to problems where experts are weighted. The paper gives several arguments supporting the claim that the weighted proportional-cumulative is perhaps the best of all Level-SP methods. Its unique weakness is the non-satisfaction of plausibility preservation.
Fortunately, it does satisfy a weaker version. Namely, if all experts agree that the probability of a level set is positive, so does society. Hence, if we are interested in problems where the final decision is level-based, then proportional-cumulative is --in our opinion-- the best IC method. In practice, the weighted proportional-cumulative can be used to aggregate experts' beliefs in various applications ranging from voting, nuclear safety, investment banking, volcanology, public health, ecology, engineering to climate change and aeronautics/aerospace. Specific examples include calculating the risks to manned spaceflight due to collision with space debris and quantifying the uncertainty of a groundwater transport model used to predict future contamination with hazardous materials (see Cooke \cite{Cooke}, Chapter 15, where those applications and others with real data are described and several aggregation methods are analyzed).

The weighted proportional-cumulative can also be combined with majority judgment to construct a robust method for electing and ranking where voters can express their uncertainties/doubts about the merit/quality of the candidates/alternatives they should judge or assess. This method is shown to avoid Arrow and Condorcet paradoxes and to resit to some natural strategic manipulations.

We see several directions of extensions. A first direction is a definition of Level-SP for a multidimensional $\Lambda$ --or more generally a metric space $\Lambda$--  and characterizing all the Level-SP methods. A second direction is a formal study of all the voting methods where voters are allowed to express their uncertainties (and compare with MJU).  

%\newpage

\section*{Acknowledgment.}
We are grateful to Thomas Boyer-Kassem, Roger Cooke, Aris Filos-Ratsikas, Herv\'e Moulin, Clemens Puppe and some anonymous referees for their helpful comments.

\newpage

The table below summarises some of the results.

\medskip

\footnotesize{
\begin{center}
\begin{tabular}{| c | c | c | }
\hline
 & General case & Anonymous case\\
\hline
Level-SP & $\sup_{S \subseteq N}  \min{(f_S(a), \min_{i\in S}\{P_i(a)\})}$ & $ median(\textbf{P(a)},f_1(a),\dots,f_{n+1}(a))$ \\
 & $lim_{a \rightarrow \sup\Lambda} f_N(a)=1$ & $lim_{a \rightarrow \sup\Lambda} f_n(a)=1$ \\
 \hline
 Certainty Preservation & $f_S$ are constants, $f_N=1$ & $f_i$ are constants,$f_{n+1}=1$ \\
 \hline
 Plausibility Preservation & $f_S$ is strictly increasing  & $f_i$ is strictly increasing \\
 & when not equal to 0 or 1 & when not equal to $0,1$\\
 \hline 
 
 Plausibility Preservation & $f_S \in \{0,1\}$ & Order Cumulatives\\
 +Certainty Preservation & & \\
 \hline
 
  Proportionality &  $\mu_{\textbf{w}}(\textbf{r}) :=\sup\left\{y | \sum_{i | r_i \geq y}w_i \geq y \right\} $&$ median(r_1,\dots,r_n,1/n,\dots,1-1/n)$\\
 \hline
\end{tabular}
\end{center}
}

\normalsize 

\appendix
\section{Introduction to the appendix.}

Our main problem is to determine a PAF (probability aggregation function) $\psi : \mathcal{P}^N \rightarrow \mathcal{P} $ satisfying some desirable properties where $\Lambda$ is a Borel subset of $\mathds{R}$ and $\mathcal{P}$ is the set of probability distributions over $\Lambda$. Let $M = \sup \Lambda$. For example, if $\Lambda=\mathds{R}$, then $M=+\infty$. 

To each probability $p\in \mathcal{P}$ is associated a unique cumulative distribution function (or CDF) $P=\pi (p)$ and vice-versa where $P(a)=\int_{x\leq a} p(x) dx$. Let $\mathcal{C}$ denote the set of cumulative distributions on $\Lambda$ (e.g. increasing and right continuous functions $Q: \Lambda \rightarrow [0,1]$ s.t. $\lim_{a\rightarrow M} Q (a)=1$ ).

Our probability aggregation problem becomes equivalent to an aggregation of CDFs. More precisely, to each PAF $\psi$ is associated a unique cumulative aggregation function (CAF)  $\Psi := \pi \circ \psi: \mathcal{C}^N\rightarrow \mathcal{C}$ where $\Psi(P_1,...,P_1)=\psi(\pi^{-1}(P_1),...,\pi^{-1}(P_n))$ and vice-versa.

We could have defined our problem directly as a cumulative aggregation problem, but it is more convenient to define and use both formulations because: 1) the existing literature is only concerned with the probability aggregation one, 2) all our axioms (except one) are more naturally formulated in the probability space, 3) the level-SP axiom and all our characterizations are more naturally formulated in the cumulative space. 

This supplementary material is organized as follows. Section \ref{Main proofs} explores the main implications of Level-SP when combined or not with the certainty preservation and/or the plausibility axioms. Section \ref{proof of the theorem general} provides a step-by-step proof for the main characterizations of level-SP methods with and without anonymity. Section \ref{certainty appendix} (resp. \ref{plausibility preservation}) establishes the characterizations when one supposes, in addition to level-SP, certainty preservation (resp. plausibility preservation). Section \ref{combining appendix} derives some additional characterizations when both certainty and plausibility preservations hold. Section \ref{appendix weighted experts} establishes the proofs related to weighted experts. Section \ref{appendix weighted} provides the rigorous proofs for the characterizations of level-SP methods that are proportional. Section \ref{appendix impossibility} proves some impossibility theorems and sectiom \ref{lambda = 3} examines the limits of the previous impossibility results.

\section{Level-SP, certainty and plausibility characterizations.}\label{Main proofs}

\subsection{Level-SP characterizations.} \label{proof of the theorem general}

%As one may , Level-SP would be better expressed in terms of cumulative functions.

If a PAF $\psi$ is level-SP, then its associated CAF $\Psi$ verifies the following property.

\begin{lemma}[level-SP, rewritten in terms of cumulative]
Suppose $\psi$ is level-SP. Then, for any expert $i$, for any level $a \in \Lambda$ and any cumulative input votes $P_1,\dots,P_n,P'_i \in \mathcal{C}$ we have:
$$ P_i(a) < \Psi(\textbf{P})(a) \Rightarrow \Psi(\textbf{P})(a) \leq \Psi(\textbf{P}_{-i}(P'_i))(a)$$
and
$$ P_i(a) > \Psi(\textbf{P})(a) \Rightarrow \Psi(\textbf{P})(a) \geq \Psi(\textbf{P}_{-i}(P'_i))(a)$$
where $\Psi := \pi \circ \psi$.
\end{lemma}

We will now establish some lemmas pertinent to level-SP that are useful in proving the main characterizations of the paper.

\begin{lemma}[Level-SP $\implies$ Monotonicity]\label{monotonicity}
If $\Psi$ is level-SP then for all experts $i$, for all levels $a\in \Lambda$, and all cumulative votes $P_1,\dots,P_n,P'_i$:

$$P_i(a) \leq P'_i(a) \Rightarrow \Psi(\textbf{P})(a) \leq \Psi(\textbf{P}_{-i}(P'_i))(a)$$
\end{lemma}

\proof{Proof:}
We will use a \emph{reductio ad absurdum}  to reach our result.
Suppose that $\Psi$ verifies level-SP but is not monotonous in all $a \in \Lambda$. Then there is a $P_i$, $P'_i$ and $a$ such that $P_i(a) \leq P'_i(a)$ and $\Psi(\textbf{P})(a) > \Psi(\textbf{P}_{-i}(P'_i))(a)$.

If $\Psi(\textbf{P}_{-i}(P'_i))(a) \geq P_i(a)$. Then:
$$\Psi(\textbf{P})(a) > \Psi(\textbf{P}_{-i}(P'_i))(a) \geq P_i(a)$$

This contradicts level-SP.

If $\Psi(\textbf{P})(a) \leq P'_i(a)$. Then:
$$P'_i(a) \geq \Psi(\textbf{P})(a) > \Psi(\textbf{P}_{-i}(P'_i))(a)$$

Then replacing $P'_i(a)$ by $P_i(a)$ improves the output, this contradicts level-SP.

Else:
$$\Psi(\textbf{P}_{-i}(P'_i))(a) < P_i(a) \leq P'_i(a) < \Psi(\textbf{P})(a)$$

This also contradicts level-SP. Therefore level-SP implies monotonicity. 
\endproof

\begin{lemma}[Level-SP $\implies$ Level Independence]\label{Level}
If $\Psi$ is level-SP then for all $a\in \Lambda$ the value of $\Psi(P_1,\dots,P_n)(a)$ only depends on $P_1(a),\dots,P_n(a)$.
\end{lemma}

\proof{Proof:}
For any $a\in \Lambda$, suppose we have $P_1,\dots,P_n$ and $P'_1,\dots,P'_n$ such that for all experts $i$, $P_i(a) = P'_i(a)$. Then according to the monotonicity lemma (Lemma \ref{monotonicity}) $\Psi(P_1,\dots,P_n)(a) = \Psi(P'_1,\dots,P'_n)(a)$. Therefore $\Psi(P_1,\dots,P_n)(a)$ only depends of $P_1(a),\dots,P_n(a)$. 
\endproof

\proof{Proof of theorem \ref{charac general}:}
$\Rightarrow:$ 
\begin{itemize}
    \item Let us show the existence of the phantom functions $f_S$.
    
The level lemma (Lemma \ref{Level}) gives us that if $\Psi$ is level-SP then for all $a \in \Lambda$ we have that the value $\Psi(P_1,\dots,P_n)(a)$ only depends on $P_1(a),\dots,P_n(a)$ as such level-SP implies the existence of a one-SP function $g_a$ such that:
\[\forall \textbf{P}, \Psi(\textbf{P})(a) = g_a(P_1(a),\dots,P_n(a)).\]
Since $g_a$ is one-SP, by Moulin \cite{M1980} classical result, we have that there are phantom values $b_S^a : S \subseteq N$ that are increasing with $S$ such that:
\[\forall \textbf{r}, g_a(\textbf{r}) =\max_{S \subseteq N}  \min{(b_S^a, \min_{i\in S}\{r_i\})}  \]
It follows that there are $2^n$ phantom functions $f_S : \Lambda \rightarrow [0,1]$  that are increasing over the subsets of $S$ such that:
\[\forall a \in \Lambda, \forall \textbf{P}, \Psi(\textbf{P})(a)  =\max_{S \subseteq N}  \min{(f_S(a), \min_{i\in S}\{P_i(a)\})} \].

    \item Let us now show that phantom functions $f_S$ are right continuous and increasing.

Let us consider any phantom function $f_S$ and any two alternatives $a < b$. Let us suppose that all experts $i \in S$ are single-minded in $a$ (every vote $p_i$ with $i \in S$ is equal to a Dirac mass at $a\in \Lambda$, e.g. $p_i=\delta_{a}, \forall i\in S$) and that all other experts are single-minded in $b$. Then the outcome of the interval $[a,b[$ is given by $f_S$. Since the outcome is a cumulative distribution it must be right continuous and increasing. As such $f_S$ is right continuous and increasing on the interval $[a,b[$. Since this is true for all intervals $[a,b[$ we have that $f_S$ is right continuous and increasing on $\Lambda$, and this is true for every $S$.

    \item Let us now show that $\lim_{a \rightarrow \sup\Lambda} f_N(a) =1.$
    
Since when inputs are cumulative distributions the outcome is a cumulative distribution we have:
\[\lim_{a \rightarrow \sup\Lambda} \Psi(\textbf{P})(a) =1, \] therefore since for all $i$, $\lim_{a \rightarrow \sup\Lambda} P_i(a)$ =1, it follows that $\lim_{a \rightarrow \sup\Lambda}  \Psi_a(\textbf{1}) =\lim_{a \rightarrow \sup\Lambda} f_N(a) =1$. Therefore:
\[\lim_{a \rightarrow \sup\Lambda} f_N(a) =1. \]
\item The proof for if $\inf \Lambda \not \in \Lambda$ then $\lim_{a \rightarrow\inf\Lambda}f_N(a) = 0$ is symetrical to the one for $\lim_{a \rightarrow \sup\Lambda} f_N(a) =1$.
\end{itemize}

$\Leftarrow:$ Let us fix $a$, let $g_a$ be the one-SP voting rule defined as:
\[\forall \textbf{r}, g_a(\textbf{r}) =\max_{S \subseteq N}  \min{(f_S(a), \min_{i\in S}\{r_i\})}.\]

Then by definition we have $\forall \textbf{P}, g_a(P_1(a),\dots,P_n(a)) = \Psi(\textbf{P})(a)$.
Since $g_a$ is a one-SP rule, we have for all alternatives $a \in \Lambda$ and all profiles $\textbf{P}$:

\[
    P_i(a) < \Psi(\textbf{P})(a)  \Rightarrow \Psi(\textbf{P})(a) \leq \Psi(\textbf{P}_{-i}(P'_i))(a)
\]
and 
\[
    P_i(a) > \Psi(\textbf{P})(a)  \Rightarrow \Psi(P)(a) \geq \Psi(\textbf{P}_{-i}(P'_i))(a).
\]

In other words, $\Psi$ is level-SP.

The increasing right continuity of the $f_S$ functions imply that the outcome is right continuous and increasing. Since $\lim_{a \rightarrow \sup \Lambda} f_N(a) = 1$, we know that $\lim_{a \rightarrow \sup\Lambda}, \Psi(\textbf{P})(a) =1$ by using the formula for the value $S$. As such, if the inputs are cumulative distributions the outcome is also a cumulative distribution.

$\circ:$ Unanimity is immediate by the condition for a unanimous one-SP function in the Moulin's max-min characterization. 
\endproof

\proof{Proof of theorem \ref{anonymous charac}:}
The proof is essentially the same as the general case except one uses the anonymous characterization. The following only gives the details not given in the general proof.

$\Rightarrow:$ Once we have shown the existence of $g_a$ we must show that if $\psi$ is anonymous then for all $a< \Lambda$ we have that $g_a$ is anonymous. Let us do this by \textit{reductio ad absurdum}. Suppose that $g_a$ is not anonymous then there is a permutation $\sigma$ and inputs $\textbf{r}$ such that $g_a(\textbf{r})\neq g_a(r_{\sigma(1)},\dots,r_{\sigma(n)})$. It follows that for any $\textbf{P}$ such that $\forall i, r_i = P_i(a)$, we have:
\[\Psi(\textbf{P})(a)\neq\Psi(P_{\sigma(1)},\dots,P_{\sigma(n)})(a).\]
As such $\Psi$ is not anonymous. It follows (by the bijection between the CDFs and the PDFs) that $\psi$ is not anonymous. QED.

$\Leftarrow:$ It remains to show that $\psi$ is anonymous. It is immediate by the characterization that all $g_a$ are anonymous. As such:
\[\forall a, \forall \textbf{P}, \Psi(\textbf{P})(a) = \Psi(P_{\sigma(1)},\dots,P_{\sigma(n)})(a).\]
Therefore:
\[\forall \textbf{P}, \Psi(\textbf{P})=\Psi(P_{\sigma(1)},\dots,P_{\sigma(n)}).\]
Hence $\Psi$ is anonymous, consequently, by the bijection between the CDFs and the PDFs, so is $\psi$. 
\endproof

\subsection{Consistency of phantom function definitions}\label{proof consistent phantoms}
%begin{lemma}
%If $\psi$ is anonymous then the two characterizations and definitions of phantom function (via max-min and median formula) are consistent and linked as follows:
%\[\forall S \subset N: f_S = f_{\# S}.\]
%where $\# S$ denote the cardinal of $S$, the ``$f$'' on the left side denotes a phantom function in the max-min formula (general case) and the ``$f$'' in the right side it denotes a phantom function in the median formula (anonymous case).
%\end{lemma}

%Thus the terminology ``phantom functions'' is not ambiguous since, in both characterizations, they represent the ``same functions.''

\proof{Proof of remark \ref{remark consistent phantoms}:}
Let $\psi$ be a level-SP PAF and let $\{f_S : S \in N \}$ be the associated phantom functions.
Let us suppose that $\forall k,  \exists h_k \in \Lambda \rightarrow [0,1], \forall S \in N : k = \#S \Rightarrow f_S = h_k$. We wish to show that $\psi$ is anonymous and that, in the anonymous characterizations, the phantom functions are $h_1,\dots, h_n$.

Let us first show that $\psi$ is anonymous. Let $\sigma$ be a permutation of $n$ elements.
\begin{align*}
    \forall a,\Psi(P_{\sigma(1)},\dots,P_{\sigma(n)})(a) &= \max_{S \subseteq N}  \min{(f_S(a), \min_{i\in S}\{r_{\sigma(i)}\})} \\
    &= \max_{k \leq n} \min{(h_k(a), \min_{i\in S : \#S = k}\{r_{\sigma(i)}\})} \\
    &= \max_{k \leq n} \min{(h_k(a), \min_{i\in S : \#S = k}\{r_i\})} \\
    &= \max_{S \subseteq N}  \min{(f_S(a), \min_{i\in S}\{r_i\})} \\
    &= \Psi(\textbf{P})(a)
\end{align*}

Therefore $\psi$ is anonymous. It follows that there exists a set of $f_k$ typed phantom functions in the anonymous characterization. Let us now show that that $h_k = f_k$.

For any given $S \subseteq n$ of size $k$. Suppose that for any given $a < b \leq \sup \Lambda$, the set $S$ of experts are single-minded in $a$ and the rest are single-minded in $b$. It follows that according to the characterization in theorem \ref{charac general}, we have $\Psi(\textbf{P})(a) = h_k(a)$ and according to the characterisation in lemma \ref{anonymous charac} we have $\Psi(\textbf{P})(a) = f_k(a)$.

Since no matter the $a\in\Lambda$ this is true, we have $f_k =h_k = f_S$. 
\endproof

\subsection{Certainty preservation.}\label{certainty appendix}

\proof{Proof of proposition \ref{certainty pres}:}

$\Rightarrow:$ Let us show that for a fixed $S \subset N$ we have that $f_S$ is a constant. Consider any two alternatives $a < b < c \leq \sup \Lambda$, and suppose that all experts in $S$ are single-minded in $a$ and the rest are single-minded in $c$. By certainty preservation we have that:
\[f_S(a)=\Psi(\textbf{P})(a)=\Psi(\textbf{P})(b)=f_S(b).\]
Since $a$ and $b$ can be arbitrarily chosen, the phantom functions are constants.

$\Leftarrow:$ Suppose the $f_S$ are all constants. For any interval $[a,b]$, suppose that for all experts $i$ we have $P_i(a) = P_i(b)$. We therefore obtain $\Psi(\textbf{P})(a) = \Psi(\textbf{P})(b)$. Therefore $\psi$ satisfies certainty preservation. 
\endproof

\subsection{Plausibility Preservation.}\label{plausibility preservation}

\proof{Proof of proposition \ref{plaus charac}:}

$\Rightarrow:$ Let us consider the phantom function $f_S$ and two alternatives $a < b < \sup \Lambda$ such that $f_S$ never takes the values $0$ or $1$ on the interval $[a,b[$. Suppose that all experts $i$ that are part of $S$ submit:
\[P_i(c)=f_S(c) +\frac{c-a}{b-a}.\]
All other experts $i$ submit:
\[P_i(c)=f_S(c)\frac{c-a}{b-a}.\]
Since phantom functions are non-decreasing and not equal to $0$, the inputs for all experts are strictly increasing on the interval $[a,b[$. Since $\psi$ is plausibility preserving we therefore have that $\Psi(\textbf{P})$ must be strictly increasing on the interval $[a,b[$. Due to theorem \ref{charac general} we know that on the interval $[a,b[$ we have $\Psi(\textbf{P}) =f_S$. Therefore we have shown that $f_S$ is strictly increasing on the interval $[a,b[$. Since this proof holds for any interval $[a,b[$ chosen where $f_S$ is never worth $0$ or $1$ we have shown that $f_S$ is strictly increasing when not worth $0$ or $1$.

The remainder of the proof uses \textit{reducto ad absurdm}. Suppose that there is $a <\sup \Lambda$ such that $f_\emptyset(a) = 1$. Then if all experts agree that $P_i(a) < 1$, then according to plausibility preservation:
\[1 =\Psi(a) < \Psi(\sup\Lambda) =1\]
This shows that our hypothesis was absurd. Therefore we can conclude that for all $a < \sup \Lambda$ $f_\emptyset(a) < 1$.

Similarly, if $f_N(a) =0$ for any $a$ then assuming all experts are single-minded in $a$. We have $p_i([a,a]) =1$ and $\psi(\textbf{p}) =0$. This contradicts plausibility preservation. It follows that $f_N > 0$.

$\Leftarrow:$ Suppose that all the phantom functions $(f_S)_{ S\subset N}$ are strictly increasing when not in $\{0,1\}$, and that $f_\emptyset < 1$ (except maybe in $\sup\Lambda$) and $f_N > 0$. For any interval $[a,b]$, suppose that for all $i$, $p_i([a,b]) > 0$.
As such for all experts $i$, we have $P_i(a) < P_i(b)$ and for all $S \subseteq N$, $f_S(a) \leq f_S(b)$ with equality iff $f_S(a) = 1$ or $f_S(b)=0$.

We will now show by \textit{reducto ad absurdm} we have plausibility preservation. Hence, suppose that:
\begin{equation}\label{equation absurd plausibility}
    \Psi(\textbf{P})(a)=\max_{S \subseteq N}  \min{(f_S(a), \min_{i\in S}\{P_i(a)\})} =
\max_{S \subseteq N}  \min{(f_S(b), \min_{i\in S}\{P_i(b)\})}=\Psi(\textbf{P})(b).
\end{equation}

There is a $T$ such that:
\[\max_{S \subseteq N}  \min{(f_S(a), \min_{i\in S}\{P_i(a)\})} = \min{(f_T(a), \min_{i\in T}\{P_i(a)\})}\]

Let us now consider $\min{(f_T(b), \min_{i\in T}\{P_i(b)\})}$. We have the following inequalities:

\[\Psi(\textbf{P})(a) =\min{(f_T(a), \min_{i\in T}\{P_i(a)\})} \leq \min{(f_T(b), \min_{i\in T}\{P_i(b)\})} \leq \max_{S \subseteq N}  \min{(f_S(b), \min_{i\in S}\{P_i(b)\})} = \Psi(\textbf{P})(b).\]

As such we have according to the equation \ref{equation absurd plausibility}:
\begin{equation}\label{equality absurd}
    \min{(f_T(a), \min_{i\in T}\{P_i(a)\})} = \min{(f_T(b), \min_{i\in T}\{P_i(b)\})}.
\end{equation}

If there is an expert $j$ such that $\min{(f_T(b), \min_{i\in T}\{P_i(b)\})} =P_j(b)$. Then since $P_j(a) < P_j(b)$, we cannot have the equality in equation (\ref{equality absurd}). As such, we have $\Psi(\textbf{P})(b) = f_T(b) < \min_{i\in T}\{P_i(b)\}$. It follows that $f_T(b) = \Psi(\textbf{P})(a) \leq f_T(a)$. Therefore $f_T(a) = f_T(b)$. Since $f_T$ is strictly increasing when not equal to $0$ or $1$, we obtain that either $f_T(a) = 1$ or $f_T(b) = 0$. 

If $f_T(a) = 1$ we contradict that $f_T(b) < \min_{i\in T}\{P_i(b)\}$. If $f_T(b) =0$, then $f_T(a)= f_T(b) < min(f_N(b),\min_{i\in N}\{P_i(b)\})$ which contradicts that $T$ is the set such that:
\[\max_{S \subseteq N}  \min{(f_S(a), \min_{i\in S}\{P_i(a)\})} = \min{(f_T(a), \min_{i\in T}\{P_i(a)\})}\]

As such we have reached a contradiction. It was absurd to assume that $\Psi(\textbf{P})(a) = \Psi(\textbf{P})(b)$. We have shown that we are plausibility preserving.  
\endproof

\subsection{Combining level-SP, certainty preservation and plausibility preservation.}\label{combining appendix}

\proof{Proof of proposition \ref{Combi charac 1}:}\label{cer + plaus proof}
$\Rightarrow: $ Since $\psi$ is level-SP and certainty preserving the phantom functions $f_S$ are constants. Since $\psi$ is level-SP and plausibility preserving the phantom functions $f_S$ are strictly increasing when not worth $0$ or $1$. As such the phantoms functions $f_S$ are constants in $\{0,1\}$ with $f_\emptyset = 0$ and $f_N =1$.

Let $g$ denote the one-SP associated voting rule and consider any input $\textbf{r}$. We will use \textit{reducto ad absurdum}. Suppose that $g(\textbf{r}) \not \in \{r_i : i \in N \}$. Then, according to theorem \ref{charac general}, we have that $g(\textbf{r}) \in \{0,1\}$.

\begin{itemize}
    \item If $g(\textbf{r}) = 1$, then according to theorem \ref{charac general} there is an $S$ such that $f_S =1$. If $ S \neq\emptyset$ then for all $i \in S$, $r_i = 1$. Since by hypothesis, $1 = g(\textbf{r}) \not \in \{r_i : i \in N \}$, none of the $r_i$ are equal to $1$. As such $S = \emptyset$. However, since $f_\emptyset = 0$ (certainty preservation) this is absurd. W have shown that we cannot have $g(\textbf{r}) = 1$.
    \item If $g(\textbf{r}) = 0$, then according to theorem \ref{charac general} $min(f_N,min_{i \in N}\{r_i\}) =0$. Since $f_N=1$ (certainty preservation) we therefore have that one of the $r_i$ is equal to $ 0$. However by hypothesis $0 = g(\textbf{r}) \not \in \{r_i : i \in N \}$. We have reached a contradiction.
\end{itemize}

As such our hypothesis was wrong and so:
\[\forall \textbf{r}, g(\textbf{r}) \in \{r_i : i \in N \}.\]

$\Leftarrow: $ Suppose that $\psi$ has an associated one-SP voting rule $g$ such that for all $\textbf{r}$, $g(r_1,...,r_n) \subset \{r_1,...,r_n\}$. Then $\psi$ is certainty preserving and level-SP. We also have $\forall \textbf{r} \in \{0,1\}^N, g(\textbf{r}) \in \{0,1\}.$ As such all phantoms functions are in $\{0,1\}$. Therefore $\psi$ is plausibility preserving. 
\endproof

\proof{Proof of proposition \ref{Combi charac 2}:}
Let $r_i$ be selected. Then according to lemma \ref{moulin g} $f_{j : r_j \geq r_i}=1$ and $f_{j : r_j> r_i} =0$. 

Suppose that there is $k$ such that $r_i \neq r_k$ and $f_{j : r_j \geq r_k}=1$ and $f_{j : r_j> r_k} =0$. Then we reach a contradiction since $f_S$ is increasing with $S$ and that either $\{j : r_j \geq r_k\} \subseteq \{i : r_j > r_i\}$ or $\{j : r_j \geq r_i\} \subseteq \{i : r_j > r_k\}$. 

It follows that $f_{j : r_j \geq r_i}=1$ and $f_{j : r_j> r_i} =0$, iff $r_i$ is the outcome. 

Since the profile is dominating we have a permutation $\sigma$ such that $P_{\sigma(1)} \leq \dots \leq P_{\sigma(n)}$.

Therefore since $f_S$ is increasing with $S$ and that $f_\emptyset=0$ and $f_N =1$ we have the existance of $i$ such that $f_{j : P_j \geq P_i}=1$ and $f_{j : P_j> P_i} =0$.

A such no matter the alternative $a$ we have $g(\textbf{P(a)})=P_i(a)$, therefore $\psi(\textbf{p}) = p_i$. 
\endproof

\proof{Proof of theorem \ref{order charac}:}
$\Rightarrow:$ As shown above, all $f_S$ are in $\{0,1\}$ and since $g(\textbf{r}) = \operatorname{median}(r_1,\dots,r_n,f_0,\dots,f_{n})$ for some real values $f_0\leq...\leq f_{n}$, there is $k$ such that $0=f_1= f_2=...=f_k<f_{k+1}=...=1 $. Thus, we deduce that $g(\textbf{r})$ is always the $k$-th greatest element of $\textbf{r}$ and so $g$ is an order function.

$\Leftarrow:$ Since an order function can be written as 
$\operatorname{median}(r_1,\dots,r_n,f_0,\dots,f_{n})$ where all the phantoms are worth $0$ or $1$, the corresponding $\psi$ is anonymous, certainty preserving, plausibility preserving and level-SP. 
\endproof

\subsection{Weighted experts.}\label{appendix weighted experts}

\proof{Proof of the phantom voters formula for theorem \ref{theorem weighted experts}}
$\Rightarrow:$ For this proof we will use the notation $w(S) := \sum_{j\in S}w_i$ and $w'(S) := \sum_{j\in S}w'_i$.
\begin{itemize}
    \item First let us show that $\gamma$ exists. Suppose there exists $S,S'$ $\textbf{w}, \textbf{w'}$ such that $\dfrac{w(S)}{w(N)} = \dfrac{w'(S')}{w'(N)}$. Let us consider 2 situations. In situation A (resp. B), we have weights $\textbf{w}$ (resp. $\textbf{w'}$) all experts $i \in S$ (resp. $i \in S'$) have $P_i(a) =1$ and all other experts have $P_i(a) =0$.
Then according to w-additivity, we can replace by the situation with 2 experts $\{1_A,2_A\}$ (resp. $\{1_B,2_B\}$) with weights $w(S),w(M)-w(S)$ (resp. $w(S'),w(M')-w(S)$). We also have $P_{1_A}(a) = P_{1_B}(a) = 1$ and $P_{2_A}(a) = P_{2_B}(a) = 0$. By w-anonymity and w-additivity we can change the name of the two experts with strictly positive weights in situation B such that they coincide with those in situation A.
By w-proportionality, due to the equality, we have that the outcome of both situations are the same. It follows that  $f_S^\textbf{w}(a) = f^\textbf{w'}_{S'}(a)$. 

Hence, we have shown that when $\dfrac{w(S)}{w(N)} = \dfrac{w(S')}{w(N)}$, then we must have that $f_S^\textbf{w} = f_{S'}^\textbf{w'}$. We can therefore find a function $\gamma$ such that for all $S$ and $M$ $f_S^\textbf{w}(a) = \gamma\left(a,\frac{w(S)}{w(N)} \right)$.
\item Let us now show that $\gamma$ is none-decreasing in its second argument. Suppose this is false. Then there are $S,S',\textbf{w},\textbf{w'}$ and $a\in \Lambda$ such that $\dfrac{w(S)}{w(N)} < \dfrac{w'(S')}{w'(N)}$ and $\gamma\left(a,\dfrac{w(S)}{w(N)}\right) > \gamma\left(a,\dfrac{w'(S')}{w'(N)}\right)$. Let us consider two situations A and B. In situation A (resp. B) we have weights $\textbf{w}$ (resp. $\textbf{w'}$) and all experts $i \in S$ (resp $i \in S'$) have $P_i(a) =1$ and all other experts have $P_i(a) =0$. 

By w-proportionality, multiplying all the weights in setting B by a constant, we obtain $w(N)=w'(N)$. It follows that we have $w(S) < w(S')$. By w-additivity, we can now consider the settings as having exactly 2 experts with positive weights. Then by w-anomymity and w-additivity we can rearrange the names of experts so as for the experts names with positive weights to coincide in both situations. It follows that the expert $i$ with $P_i(a)=1$ and positive weight prefers the situation B to A despite having a lower weight. This contradicts the w-monotonicity assumption. This is a contradiction. Therefore $\gamma$ is increasing in its second variable.
\item The remaining conditions are necessary to ensure that the phantom functions are none-decreasing right-continuous and that $\lim_{a \rightarrow \sup \Lambda} f_S^\textbf{w}(a) = 1$ 
\end{itemize}
$\Leftarrow:$ Suppose that $\gamma$ exists. It is immediate that w-additivity, w-proportionality and w-anonymity are satisfies. Suppose that between $\textbf{w}$ and $\textbf{w'}$ an expert's weight $i$ decreases everything else equal. For any $S$ that does not include $i$ we have $f_S^\textbf{w} \leq f_S^\textbf{w}  \leq f_{S+i}^\textbf{w'}  \leq f_{S+i}^\textbf{w'}$. It follows that the expert cannot benefit from this change. Therefore, the WPAF is w-monotone.  
\endproof

\proof{Proof of the grading curve formula for theorem \ref{theorem weighted experts}}
 For a fixed $a\in \Lambda$, we define $\lambda_a$ and $\mu_a$ as follows:
\[\forall \textbf{w},\forall \textbf{r}, \lambda_a(\textbf{w},\textbf{r}) := \sup \left\{ y | \gamma\left(a,\dfrac{\sum_{i: r_i \geq y}w_i}{\sum_{i \in N}w_i}\right) \geq y \right\}.\] 

\[\forall \textbf{w},\forall \textbf{r}, \mu_a(\textbf{w},\textbf{r}) := \max_{S \subseteq N} \min\left(\gamma\left(a,\dfrac{\sum_S w_i}{\sum_N w_i}\right), \min\{r_i: i \in S\}\right)\]

Since we have already shown that $\mu_a$ is the One-SP voting rule that provides the characterization at level $a$, it remains to show that $\mu_a = \lambda_a$. 

Let us fix $\textbf{r}\in [0,1]^N$.
\begin{itemize}
    \item Suppose that there is $j$ such that $\mu_a(\textbf{r}) =r_j$. Then there is $S$ such that $j = \arg\min\{r_i : i \in S\}$, $f_{S-i}^\textbf{w}(a) \leq r_j \leq f_S^w(a) = \gamma\left(a,\dfrac{w(S)}{w(N)}\right)$ and if $i \not \in S$ then $r_i \leq r_j$. Let us consider $y =r_j$. Then we have :
    \[\gamma\left(a,\dfrac{\sum_{i: r_i \geq r_j}w_i}{\sum_{i \in N}w_i}\right) = f_S^\textbf{w}(a) \geq r_j. \]
    Therefore $\lambda_a(\textbf{r}) \geq r_j$.

    Let us now consider any $\epsilon>0$. Let $y=r_j + \epsilon$, we have:
    \[\gamma\left(a,\dfrac{\sum_{i: r_i \geq r_j+\epsilon}w_i}{\sum_{i \in N}w_i}\right) =f_{S-i}^\textbf{w}(a) \leq r_j < r_j + \epsilon.\]

    Therefore, $\lambda_a(\textbf{r}) \leq r_j$. Consequently, $\lambda_a(\textbf{r}) =r_j$.
    \item Let us now consider the case where no $r_i$ was selected; as such there is $S\subseteq N$ such that $\mu_a(\textbf{r})=f^\textbf{w}_S(a)$. We deduce that $r_i > f^\textbf{w}_S(a)$ iff $i \in S$.

    Let us consider the case $y=f_S^\textbf{w}(a)$. Then:
    \[\gamma\left(a,\dfrac{\sum_{i: r_i \geq f_S^\textbf{w}}w_i}{\sum_{i \in N}w_i}\right) = f_S^\textbf{w}(a).\]
    As such $\lambda_a(\textbf{r}) \geq f_S^\textbf{w}(a)$.

    Finally, let us consider, for any $\epsilon>0$, the case where $y=f_S^\textbf{w}(a)+\epsilon$. Then:
    \[\gamma\left(a,\dfrac{\sum_{i: r_i \geq f_S^\textbf{w}+ \epsilon}w_i}{\sum_{i \in N}w_i}\right) \leq  f_S^\textbf{w}(a) <  f_S^\textbf{w}(a) + \epsilon.\]
    As such $\lambda_a(\textbf{r}) \leq f_S^\textbf{w}(a)$.
  Consequently, $\lambda_a(\textbf{r}) = f_S^\textbf{w}(a)$.
\end{itemize} 
We have shown the two function are equal. As such, we can characterize using either one.

\endproof

\section{Characterization of the weighted proportional cumulative.}\label{appendix weighted}

\subsection{Main weighted theorem.}\label{appendix main weighted}

That weighted proportionality axiom can be rewritten in terms of CAF as follows:
\[\forall (a_1,\dots,a_n) \in \Lambda^N, \Psi(\delta_{a \geq a_1},\dots,\delta_{a \geq a_n} )(a)=\sum_{i : a_i\leq a} w_i. \]

This CAF formulation is useful in the proofs but the PAF formulation is more elegant and intuitive. 

\proof{Proof of theorem \ref{theorem-proportional-cumulative}:}\label{poof weight}
$\Rightarrow:$ Let $\psi$ be level-SP and weighted proportional. Let $\Psi$ be the cumulative aggregation function associated to $\psi$ and $f_S$ be its phantom functions.
Let us first show that:
\[ \forall S \subseteq N,\forall a \in \Lambda: f_S(a) = \sum_{i \in S} w_i.\]
Let us take any $S \subseteq N$ and $a < \sup\Lambda$. Suppose that all experts $i \in S$ are single-minded in $a$ and all other experts are single-minded in any $b>a$. By weighted proportional we therefore have
$g_a(P_1(a),\dots,P_N(a)) = \sum_{i \in S} w_i$. By level-SP, we have $g_a(P_1(a),\dots,P_N(a)) =f_S(a)$. We conclude that for all alternatives $a < \sup\Lambda$ we have $f_S(a) = \sum_{i \in S} w_i$ and we can assume without loss of generality that this also holds for $a = \sup\Lambda$. Hence we conclude that the phantoms associated to $\psi$ are the constant phantom functions $f_S = \sum_{i \in S} w_i$. Consequently, $\psi$ is certainty preserving, and we denote by $g$ the associated voting rule. We want to show that $g=\mu_w$.

\begin{itemize}
    \item Suppose that there is $j$ such that $g(\textbf{r}) = r_j$, then there is $S$ such that $r_j = min \{r_k : k \in S \}$ and $r_j \leq f_S$ (theorem \ref{charac general}). Let $S' =  \{k : r_j < r_k \}$. By the same theorem we have that $f_{S'} \leq r_j$.
\begin{itemize}
    \item For any $y > r_j$, we have $\sum_{i : r_i \geq y} w_i \leq \sum_{i : r_i > r_j} w_i = f_{S'} \leq r_j < y$. Therefore by definition of $\mu_w$ we have $\mu_w(\textbf{r}) \leq r_j$. 
    \item For any $y \leq r_j$, we have $\sum_{i : r_i \geq y} w_i \geq \sum_{i : r_i \geq r_j} \geq \sum_{i : r_i \in S} = f_S \geq r_j \geq y$. Therefore by definition of $\mu_w$, $\mu_w(\textbf{r}) \geq r_j$.
\end{itemize}
By combining the above we obtain that if there is $j$ such that $g(\textbf{r}) = r_j$, then 
 $\mu_w(\textbf{r}) = r_j$.
    \item Suppose that there is no $i$ such that $g(\textbf{r}) = r_i$, then by theorem \ref{charac general} there is an $S$ such that $g(\textbf{r}) =f_S$ and $f_S \leq \min_{i \in S} r_i$ therefore if $k \in S$ then $r_k > f_S$. Suppose there is an $i$ such that if $k \not \in S$ and $r_k > f_S$ then for $S' = S \cup r_k$ since $f_S' \geq f_S$ we have $ \min(f_{S'},\min_{i \in S'} r_i ) \geq f_S$. Therefore according to the theorem, $f_S = f_{S'}$. It follows that we can choose $S$ such that for all $k \not \in S$, $r_k < f_S$.
    \begin{itemize}
        \item For any $y > f_S$ we have $\sum_{i : r_i \geq y} w_i \leq \sum_{i : r_i \geq f_S} w_i = f_S < y$, therefore $\mu_w(\textbf{r}) \leq r_j$. 
        \item For $y \leq r_j$, we have $\sum_{i :  r_i \geq y} w_i \geq \sum_{i : r_i \geq f_S} w_i = f_S \geq y$ therefore $\mu_w(\textbf{r}) \geq f_S$. Therefore $\mu_w(\textbf{r}) = f_S$.
    \end{itemize}
It follows that $g= \mu_w$.
\end{itemize}

$\Leftarrow:$ If $\Psi$ is the certainty preserving, level-SP CAF associated with $\mu_w$ then for $\textbf{r} \in \{0,1\}^N$, we have $\mu_w(\textbf{r}) = \sum_{i :  r_i = 1} w_i$. As such $\psi$ verifies weighted proportionality. 
\endproof

\subsection{Rational weights} \label{appendix rational median}
\proof{Proof of proposition \ref{Prop_rational_weights}:}
First let us consider the anonymous case (all weights are equal) and suppose that we have $d$ experts. According to the weighted proportionality property $f_{\#S} = f_S = \frac{\#S}{d}$. As such:
\[\forall \textbf{t}=(t_1,...,t_d), \mu_d(\textbf{t}):=\mu_{\textbf{w}=(\frac{1}{d},...,\frac{1}{d})}(\textbf{t}) :=\operatorname{median}(t_1,\dots,t_d,0,1/d,\dots,1-1/d,1).\]

Now that we have established the characterization of $\mu_d$, we wish to establish that if we have $n$ experts such that each expert has weight $s_i/d$ where $s_i \in \mathds{N}$, then: 

\begin{equation}\label{equation duplicate}
    \forall \textbf{r}, \mu_w(\textbf{r}) = \mu_d(\overbrace{r_1,\dots,r_1}^{s_1},\dots,\overbrace{r_n,\dots,r_n}^{s_n}) = \operatorname{median}(\overbrace{r_1,\dots,r_1}^{s_1},\dots,\overbrace{r_n,\dots,r_n}^{s_n},1/d,\dots,1 -1/d).
\end{equation}

Intuitively, we simply consider that for each expert $i$, if expert $i$ has weight $w_i = s_i/d$ then they are duplicated so as to appear $s_i$ times in the anonymous case with $d$ players.

Let us now provide a formal proof. We will show that
\[\forall \textbf{r}, \lambda_\textbf{w}(\textbf{r}) = \operatorname{median}(\overbrace{r_1,\dots,r_1}^{s_1},\dots,\overbrace{r_n,\dots,r_n}^{s_n},1/d,\dots,1 -1/d).\]
verifies the weighted property. The uniticity of a level-SP function that verifies the weighted property will then allow us to conclude that $\lambda_\textbf{w} = \mu_\textbf{w}$.

Let us consider \textbf{r} where all the values are $0$ or $1$. Let $S$ be the set of experts $i$ such that $r_i=1$.
Since we have $s_i$ iterations of $r_i$, the number of iterations of $1$ in the equation defining $\lambda_\textbf{w}$ is $\sum_{i\in S} s_i$. Conversely the number of iterations of $0$ is $\sum_{i\not\in S} s_i = d - \sum_{i \in S} s_i$.
It follows that the outcome of $\lambda_\textbf{w}$ is the $k_{th}$ phantom value where $k = \left(\sum_{i \in S} s_i\right)$. However, $f_k=\frac{k}{d} = \frac{\sum_{i \in S} s_i}{d}  =\sum_{i \in S} w_i$. This shows that $\lambda_\textbf{w}$ verifies the weighted property. QED.  
\endproof
 
 \subsection{Dominated opinions.}\label{appendix weight dominated}

\proof{Proof of proposition \ref{proposition dominated prop}:}
According to proposition \ref{theorem-proportional-cumulative} we have:
\[\forall a, \forall \textbf{P}, \Psi(\textbf{P})(a) = \sup\left\{y \middle| \sum_{i : P_i(a) \geq y}w_i \geq y \right\}.\]
Suppose for all $i\in N$, $P_i \geq P_{i+1}$. Wlog we will also suppose that all the weights are strictly positive. Let us determine the value of $\Psi(\textbf{P})(a)$ for any given $a$.

\begin{itemize}
    \item Suppose $\sum_{k\leq i-1} w_k \leq P_i(a) < \sum_{k\leq i} w_k$.
    \begin{itemize}
        \item For any $y > P_i(a)$, we have $\{ k : P_k(a) \geq y \} \subseteq \{ k < i \}$. Therefore, $\sum_{k : P_k(a) \geq y}w_k \leq \sum_{k \leq i-1} w_k \leq P_i(a) < y$. By definition of $\mu_w$, we therefore have $\Psi(\textbf{P})(a) \leq P_i(a)$.
        \item For any $y \leq P_i(a)$, we have $\{ k \leq j \} \subseteq \{k : P_k(a) \geq y \} $. As such, $\sum_{k : P_k(a) \geq y}w_k \geq \sum_{k \leq j} w_k \geq P_i(a) \geq y$. By definition of $\mu_w$, we therefore have $\Psi(\textbf{P})(a) \geq P_i(a)$.
    \end{itemize} 
    We have therefore shown that $\Psi(\textbf{P})(a) = P_i(a)$. 
    
    By Bolzano's theorem there is a $a_1 \leq a$ such that $P_i(a_1) = \sum_{k\leq j-1} w_k$ and $a_2 > a$ such that $P_i(a_2) = \sum_{k\leq j} w_k$. Therefore on the interval $[a_1,a_2]$, $\Psi(\textbf{P}) = P_i$. It follows that $\psi(\textbf{p})(a) = p_i(a)$ on the interval $]a_1,a_2[$.
    
    \item Suppose that there is no $i$ such that $\sum_{k\leq i-1} w_k \leq P_i(a) < \sum_{k\leq i} w_k$. Then there is an $i$ such that, $P_{i+1}(a) < \sum_{k\leq i} w_k < P_i(a)$. We show that $\Psi(\textbf{P})(a) = \sum_{k\leq i} w_k$ by comparing $y$ to $\sum_{k\leq i} w_k$ just as we compared $y$ to $P_i(a)$ in the previous section
    %\begin{itemize}
    %\item For any $y$ such that $y \geq \sum_{k \leq i} w_k$ we have $\{k : P_k(a) \geq y\} \subseteq \{ k \leq i\}$ as such $\sum_{ k : P_k(a) \geq y} w_k \leq y$. It follows that $\Psi(\textbf{P})(a) \leq \sum_{k \leq i} w_k$. 
    %\item For any $y$ such that $y \leq \sum_{k \leq i} w_k$ we have $\{k \leq i\} \subseteq \{k : P_k(a) \geq y \}.$ As such $y \leq \sum_{k : y \leq P_k(a)} w_k$. It follows that $\Psi(\textbf{P})(a) \geq \sum_{k \leq i} w_k$. 
    %\end{itemize}
    %We have therefore shown that $\Psi(\textbf{P})(a) = \sum_{k\leq i} w_k$.
    
    By Bolzano's theorem there is a value $a_1 < a$ such that $P_i(a_1) = \sum_{k\leq j} w_k$ and $a_2 > a_1$ such that $ P_{i+1}(a_2) = \sum_{k\leq j} w_k$. We have $\Psi(\textbf{P})(a_1) = \Psi(\textbf{P})(a_2) = \sum_{k\leq i} w_k$.
    
    Therefore $\Psi(\textbf{P})(a_1) = \Psi(\textbf{P})(a_2)$. It follows that on the interval $]a_1,a_2[$ $\psi(\textbf{p})(a) = 0$.
\end{itemize}
Therefore, $\psi(\textbf{p})$ is either following one of the experts' distribution or is equal to 0.  
\endproof

\section{Impossibility results.}\label{appendix impossibility}

\subsection{Strategyproofness in the probability space}\label{proof budget dictatorship.}

\proof{Proof of remark \ref{impossibilty l1 centainty}:}
We will first prove our theorem for $2$ experts.

Suppose that $f_{\{1\}} \leq f_{\{2\}} <1$. We will reach a contradiction.

Let $A>0$ and $B$ be such that $ f_{\{2\}} + A < B$ and $A+B <1$.
Let $p_1 := (0,f_{\{2\}} + A,0,1-(f_{\{2\}} + A))$ and $p_2 :=(f_{\{2\}} + A,0,B-f_{\{2\}},1-A-B)$.

\begin{align*}
    g(r_1,r_2) = \operatorname{median}(r_1,r_2,f_{\{2\}}) \operatorname{if} r_2 \geq r_1 \\
     g(r_1,r_2) = \operatorname{median}(r_1,r_2,f_{\{1\}}) \operatorname{if} r_2 \leq r_1 \\
    p_1 =&(0,f_{\{2\}} + A,B-f_{\{2\}}-A,1-B) \\
    P_1 =&(0,f_{\{2\}} + A,B,1)) \\
    p_2 =& (f_{\{2\}} + A,0,B-f_{\{2\}},1-A-B) \\
    P_2 =& (f_{\{2\}} +A, f_{\{2\}}+A, B+A,1) \\
    \Psi(\textbf{P}) =& 
    (f_{\{2\}},f_{\{2\}}+A,B,1) \\
    \psi(\textbf{p}) =& (f_{\{2\}},A,B -f_{\{2\}} -A,1-B)\\
 \| \psi(\textbf{p}) -p_2 \|^\mathcal{P}_1 =& |f_{\{2\}} - (f_{\{2\}} + A) | + |A - 0| + |(B -f_{\{2\}} -A) - (B-f_{\{2\}})| \\ & + |(1-B) - (1 - B -A)| \\
\| \psi(\textbf{p}) -p_2 \|^\mathcal{P}_1 =& 4A.
\end{align*}

Let us now consider what happens when the second expert lies and submits the probability function $p'_2 = (f_1,0,f_2-f_1,1 - f_2)$.

\begin{align*}
    p_1 =&(0,f_{\{2\}} + A,B-f_{\{2\}}-A,1-B) \\
    P_1 =&(0,f_{\{2\}} + A,B,1)) \\
     p'_2 =& (f_{\{2\}},0,B-f_{\{2\}},1 - B) \\
    P'_2 =& (f_{\{2\}}, f_{\{2\}}, B,1) \\
    \Psi(\textbf{P'}) =& 
    (f_{\{2\}},f_{\{2\}},B,1) \\
    \psi(\textbf{p'}) =& (f_{\{2\}},0,B -f_{\{2\}},1-B)\\
 \| \psi(\textbf{p'}) -p_2 \|^\mathcal{P}_1 =& |f_{\{2\}} - (f_{\{2\}} + A) | + | 0 -0| \\ &+ |(B -f_{\{2\}}) - (B-f_{\{2\}})| + |(1-B) - (1 - B -A)| \\
\| \psi(\textbf{p'}) -p_2 \|^\mathcal{P}_1 =& 2A.
\end{align*}

Therefore our hypothesis is false. We therefore have that $f_{\{2\}}=1$.
A similar proof gives us that $f_{\{1\}} =0$. As such:
\[\forall \textbf{r}, r_1 \leq r_2 \Rightarrow \operatorname{median}(r_1,r_2,1)=r_2 \]
and
\[\forall \textbf{r}, r_1 \leq r_2 \Rightarrow \operatorname{median}(r_1,r_2,0)=r_2.\]

Therefore if $f_{\{1\}} \leq f_{\{2\}}$ we have that the second expert is a dictator. Else the first expert is a dictator.

It remains to show that we can add new experts.

Let us suppose that we have shown that if we have $n$ experts then one is a dictator. Let us now consider $n+1$ experts.

Let us consider a level-SP and certainty preserving $\psi$ with $n+1$ experts, let $g$ be its associated voting rule. For any expert $i$ there is a level-SP and certainty preserving $\varphi_i$ (with associated voting rule $g_i$) such that:
\[\forall \textbf{P} \in \mathcal{C}^n, \varphi_i(\textbf{P}) = \psi((\textbf{P}),P_i). \]
Since $\varphi_i$ is certainty preserving and level-SP it is a dictatorship, let $v_i$ be the dictator. As such so are all the $g_i$.

Suppose that there exists $i$ such that $v_i \neq i$. Then for any $r_i \neq r_{v_i}$ we have $g(\textbf{r},r_i) = r_{v_i}$ therefore by level-SP for any value $r_{n+1}$ we have $g(\textbf{r},r_{n+1}) = r_{v_i}$. Therefore $\psi$ is a dictatorship with dictator $v_i$.

Else for all $i \leq n$ we have that $g(\textbf{r},r_i) = r_i$. As such for $S$ such that ${n+1} \in S$ by considering $\textbf{r}$ such that for all $i \in S$ we have $r_i = r_{n+1}$ and all experts $i\not \in S$ verify $r_i=0$ we find that $f_S \geq r_{n+1}$. As such $f_S=1$. Similarly if $S$ does not have the expert $n+1$ then $f_S = 0$. This positioning of the phantoms corresponds to the dictatorship with dictator $n+1$.

We have shown that no matter what $\psi$ is a dictatorship. 
\endproof

\subsection{Strong plausibility.}\label{strong plausibility}

%Hence, the same impossibility result holds if one replaces unanimity with certainty preservation as the later implies the former under level-SP.

\proof{Proof of theorem \ref{ impossibility strong_plau}:}
Let $\psi$ be a probability aggregation function that is level-SP, unanimous, and strong plausibility preserving. This proof is done with a successions of \textit{reductio ad absurdum}.

\begin{itemize}
\item Suppose we can choose $\textbf{p}$ and experts $i \neq j$ such that there is $a<b$ that verify $\Psi(\textbf{P})(a) =P_i(a) = P_i(b) = \Psi(\textbf{P})(b)$ and for all experts $k\neq i$  we have $P_k(a) \neq \Psi(\textbf{P})(a)$ and $p_k([a,b]) > 0$. Suppose we can also have $[c,d[$ disjoint from $[a,b]$ such that $\Psi(\textbf{P})(c) =P_j(c) =P_j(d) = \Psi(\textbf{P})(d)$ and $P_i(c) \neq \Psi(\textbf{P})(c)$ and $p_i([c,d[) > 0$.

For such an example, we contradict strong preservation for $A = [a,b] \cup [c,d[$. As such no such example exists.

\item Let us show that their exists an expert $i$ and an interval $[a,b]$ such that we can find $\textbf{p}$ that verifies $\Psi(\textbf{P})(a) =P_i(a) = P_i(b) = \Psi(\textbf{P})(b)$.
By unanimity, we have $f_\emptyset = 0$ and $f_N = 1$. The previous proves that there is an expert $i\in N$ such that for all $S$ without $i$ there is a positive and finite number of alternatives $a_S$ such that there exists $S$ where $f_S(a_S) = 0$ and $f_{S\cup \{i\}}(a_S) >0$ and for all $a>a_S$ $f_S(a) > 0$. As such, since $\Lambda$ is rich we can find an interval $[a,b]$ and an $S$ without $i$ such that for all $a' \in [a,b]$ we have $f_S(a') =0$ and $f_{S\cup\{i\}}(a') > 0$. Let us choose $\textbf{P}$ such that $ 0 < P_i(a) = P_i(b) < f_{S\cup\{i\}}(a)$ and $P_i(b) < P_k(a) < P_k(b)$ if $k \in S$ and $P_k(a) < P_k(b) < P_i(a)$ if $k\not \in S$ and $k\neq i$. This profile verifies $\Psi(\textbf{P})(a) =P_i(a) = P_i(b) =\Psi(\textbf{P})(b)$.

\item As such we can find an expert $i$ and $a<b$ such that $\Psi(\textbf{P})(a) =P_i(a) = P_i(b) = \Psi(\textbf{P})(b)$ and for all experts $k\neq i$  we have $P_k(a) \neq \Psi(\textbf{P})(a)$ and $p_k([a,b]) > 0$. Therefore for all intervals $[c,d]$ disjoint from $[a,b]$ where $p_i(c,d) =0$ either (1) we cannot find an expert $j \neq i$ such that $\Psi(\textbf{P})(c) =P_j(c) =P_j(d) = \Psi(\textbf{P})(d)$ or (2) $P_i(c) = \Psi(\textbf{P})(c)$.

\item Suppose that (1) is false. If we can find an interval $[c,d[$ disjoint of $[a,b]$ such that if all experts $k \neq i$ are constant and equal on $[c,d]$ then $\Psi(c) =P_k(c) =P_k(d) = \Psi(\textbf{P})(d)$.  For all sub-intervals $[c',d'[$ of $[c,d[$ this is true. As such, since (2) is true, $P_i(c') = \Psi(\textbf{P})(c')$. Therefore $p_i([c,d])=0$, this is absurd.

\item Therefore from now on let us suppose that (1) is true. From now on, all experts $k \neq i$ have the same probability distribution, For any interval $[c,d]$ disjoint from $[a,b]$, if $P_k(c) = P_k(d)$, we either have $\Psi(\textbf{P})(c) \neq P_k(c)$ or $\Psi(\textbf{P})(d) \neq P_k(d)$.

\item Phantoms functions and cumulative distributions are right continuous. As such, for any interval $[c,d]$, there is a sub interval $[c',d']$ such that either $f_{N-i} > P_i$ or $f_{N-i} < P_i$ or $f_{N-i} = P_i$. Similarly, there s a sub interval $[c',d']$ such that either $f_i > P_i$ or $f_i < P_i$ or $f_i = P_i$

\item Suppose that $f_{N-i} > P_i$ over an interval $[c,d]$ disjoint from $[a,b]$. Then by Moulin's characterization $\Psi(\textbf{P})(c) = \min(P_k(c), f_{N-i}(c))$.
We can find $d'$ such that we can choose $P_i(d') < P_k(c) =P_k(d') < f_{N-i}(c)$ ($\Lambda$ is rich enough). Therefore $\Psi(\textbf{P})(c) =P_k(c) =P_k(d') = \Psi(\textbf{P})(d')$. This contradicts (1). Similarly if $f_{N-i} < P_i$  over an interval $[c,d]$, we contradict (1). As such we have $f_{N-i} \geq P_i \geq f_i$ over all intervals $[c,d]$ disjoint from $[a,b]$. 

\item Since $f_{N-i} \geq P_i \geq f_i$ over all intervals $[c,d]$ disjoint from $[a,b]$ by switching the roles of the two intervals by setting a distribution where $p_i([c,d])>0$ and $p_i([a,b]) =0$, we find that for all intervals $I \subseteq \Lambda$ and no matter the choice of $p_i$, we have $f_{N-i} \geq P_i \geq f_i$. Therefore we have that $f_i = 1$ and $f_{N-i}=0$.

\end{itemize}
Therefore $i$ is a dictator. 
\endproof

%When $\Lambda$ is finite or discrete, strong plausibility preservation is equivalent to plausibility preservation. Otherwise, it is too strong, as its combination with level-SP leads to impossibility.

\subsection{Weak Diversity.}

\begin{axiom}[Weak Diversity]
For every diracs inputs $(p_1,...,p_n)=(\delta_{a_1},...,\delta_{a_n})$, there exists positive weights $w_1>0,...,w_n>0$ such that $\psi(\delta_{a_1},...,\delta_{a_n})=\sum_{i=1}^n w_i \delta_{a_i}$.  
\end{axiom}

Why do we call it weak diversity? because the axiom requires that, all experts possible opinions (when they are degenerate) have a positive probability in the output. 

%(e.g. every expert input is absolutely continuous with respect to the output, when the inputs are finitely supported). 

Weighted proportional methods are weak diverse. But this axiom is much weaker than the weighted proportionality axiom because the expert's weights may depend on the input in the weak diversity axiom while in the case of proportionality, the expert's weights are the same for all inputs. This opens the door to a much larger class of methods, that can be characterized as follows.

\begin{proposition*}
The association of weak diversity with Level-SP is equivalent to certainty preservation $+$ that $f_S$ is strictly increasing with respect to $S$.
\end{proposition*}

\proof{Proof:}
$\Rightarrow:$ Suppose $\psi$ verifies weak diversity and level-SP. Let $a < M$, and suppose that all experts are single minded, that is the input is $(\delta_{a_1},...,\delta_{a_n})$.

If no expert wanted elements in the interval $]a,b]$ ($a_i \not \in ]a,b]$ for all $i$) then by weak diversity $\Psi(a) = \Psi(b)$. 
As such for any $S$ we have that if the experts $i$ in $S$ have $a_i \leq a$ and the rest have $b < a_i$ then $f_S(a) = \Psi(a) = \Psi(b)=f_S(b)$. Since we can consider any interval we can conclude that the $f_S$ are constants. Therefore we are certainty preserving. 
Since whenever all experts are single minded, no matter what alternative they choose the weight of each of those alternative must be felt ($w_i> 0$) then for all $S \neq \emptyset$ for all $i\in S$, $f_S > f_{S-i}$.

$\Leftarrow:$ Suppose certainty preservation and that $f_S$ is strictly increasing. Then, for any input of single-minded experts $(\delta_{a_1},...,\delta_{a_n})$ and all $a\in \Lambda$
\[\Psi(\textbf{P})(a) = f_{\{ i : a_i \leq a \}}(a)\]
Therefore, $\psi(\delta_{a_1},...,\delta_{a_n})(a) = f_{\{ i : a_i \leq a \}} - f_{\{ i : a_i < a \}}$ and so:
\[ \psi(\delta_{a_1},...,\delta_{a_n})=\sum_{i=1}^n (f_{\{ i : a_j \leq a_i \}} - f_{\{ i : a_j < a_i \}}) \delta_{a_i} \] 
\endproof

By construction, the weighted proportional cumulative methods satisfy weak diversity. However, we have seen that they do not preserve plausibility. On the other hand, the order-cumulative PAFs are not weak diversified but they do preserve plausibility. Are there methods that are weak diversified and plausibility preserving? Unfortunately, no!

\begin{theorem}[Weak diversity Impossibility] For 3 or more experts, there are no level-SP probability aggregation functions that satisfy weak diversity and preserve plausibility.
\end{theorem}

\proof{Proof:}\label{impossibiliy proof 3}
Suppose that $\psi:\mathcal{P}^N \rightarrow \mathcal{P}$ is a level-SP probability aggregation function that satisfies weak diversity and preserves plausibility. Then, since weak diversity implies certainty preservation and that $\psi$ preserves plausibility, we must have $f_S \in \{0,1\}^N$. On the other hand, weak diversity implies that $f_S$ is strictly increasing with $S$. By pigeon-hole principle, we reach a contradiction. 
\endproof

Unfortunately, weighted proportional cumulative methods do satisfy a weaker version of plausibility, namely, if all experts agree that a level event has positive probability, so does society. This is good enough for the applications where the regulator's final decision only depends on the probability of some level events. 

\section{Combining the two notions of SP.}\label{lambda = 3}

Here we prove the existence of methods that are at the same time strategyproof in the cumulative (level-SP) and in the probability spaces ($L_1^{\mathcal{P}}-SP$ as in \cite{EC2019}). 
\begin{proposition*}
For $\#\Lambda=3$, any Level-SP PAF is also $L_1^{\mathcal{P}}-SP$. 
\end{proposition*}

\proof{Proof:}
Suppose that $\Lambda$ has three alternatives $\{a_1,a_2,a_3\}$.

\[\|\psi(\textbf{P}) -p_i\|_1 = |\Psi(\textbf{P})(a_1) -P_i(a_1)| + |\Psi(\textbf{P})(a_2) -\Psi(\textbf{P})(a_1)  -P_i(a_2) + P_i(a_1)| + |\Psi(\textbf{P})(a_2) -P_i(a_2)|\]

Let $\textbf{Q}$ only differ from $\textbf{P}$ in dimension $i$. 

Suppose that $\|\psi(\textbf{P}) -p_i\|_1 > \|\psi(\textbf{Q}) -p_i\|_1$. 

By level-SP it is impossible to decrease $|\Psi(\textbf{P})(a_1) -P_i(a_1)|$ or $|\Psi(\textbf{P})(a_2) -P_i(a_2)|$. Therefore we decreased $|\Psi(\textbf{P})(a_2) -\Psi(\textbf{P})(a_1)  -P_i(a_2) + P_i(a_1)|$. If you did this by causing a change to the value of $\Psi(\textbf{P})(a_2)$ then you increased $|\Psi(\textbf{P})(a_2) -P_i(a_2)|$ by the same amount (resp for $\Psi(\textbf{P})(a_1)$). Therefore we cannot decrease the value for the $L_1$ norm. We have reached our contradiction. 
\endproof

The converse of the above proposition is false. The IMM method introduced in \cite{EC2019} has been shown to be proportional and $L_1^{\mathcal{P}}-SP$. The table below shows that the IMM differs from the proportional cumulative. As the proportional cumulative is the unique proportional Level-SP method, the IMM is not level-SP for 3 alternatives.\\

\begin{center}
\begin{tabular}{ | c | c | c | c | c | c | c | c | c || c |}
\hline
 & $p_1$ & $P_1$ & $p_2$ & $P_2$ & $p_3$ & $P_3$ & Prop. Cum. ($\Psi$) & Prop. Cum. (density $\psi$) & $IMM$ (density) \\
 \hline
 a & 0 & 0 & 0.5 & 0.5 & 0.9 & 0.9 & 0.5 & 0.5 & 0.4\\ 
 b & 0.5 & 0.5 & 0 & 0.5 & 0 & 0.9 & 0.5 & 0 & 0.4\\ 
 c & 0.5 & 1 & 0.5 & 1 & 0.1 & 1 & 1 & 0.5 & 0.2 \\
 \hline
\end{tabular}
\end{center}

\medskip

In the Table, the lines represent 3 ordered alternatives $a<b<c$. Column 1 is the input probability distribution $p_1$ of voter 1 (its density function), column 2 represents the cumulative distribution $P_1$ of $p_1$, and similarly for voters 2 (column 3 and 4) and voter 3 (column 5 and 6). Column 7 is the cumulative distribution of the proportional cumulative (for $x=a,b,c$, $\Psi(x)=med(P_1(x),P_2(x),P_3(x),1/3,2/3)$). Column 8 is the density function of the proportional cumulative computed in Column 7. The last column represents the density function of the IMM (its calculation is borrowed from \cite{EC2019} where this numerical example is presented).

\begin{proposition*}
There is an infinite number of $\psi$ functions that are Level-SP and $L_1$-SP for $4$ alternatives.
\end{proposition*}

But such methods cannot be certainty preserving, unless dictatorial, as shown above.

\proof{Proof}\label{example}
Let us choose a Level-SP function $\Psi$ such that, $\Psi_1 = min$ and $\Psi_3 = max$.

\begin{itemize}
    \item Let us suppose that by changing $P_1$ player $i$ can improve their output for $|\psi_j(a_2) -p_j(a_2)|$.
Then they decrease $\Psi_1$ which is detrimental for $|\psi_j(a_1) -p_j(a_1)|$. Therefore any change to $P_1$ cannot improve the $\|\circ\|_1$ distance.
    \item Let us suppose that by changing $P_3$ player $i$ can improve their output for:
$$|\psi_j(a_3) -p_j(a_3)|$$.
Then they increase $\Psi_3$ which is detrimental for $|\psi_j(a_3) -p_j(a_3)|$. Therefore changes to $P_3$ cannot improve the $\|\circ\|_1$ distance.
    \item Suppose that $p_2 > \psi_2$ then since $p_1 \geq \psi_1$ we have $P_2 > \Psi_2$. Therefore by level SP it is impossible to improve $|p_2 - \psi_2|$.
    \item Similarly if $p_3 > \psi_3$ then by level SP it is impossible to improve $|p_2 - \psi_2|$.

    \item Suppose that $p_2 \leq \psi_2$ and $p_3 \leq \psi_3$. Therefore any changes to $\Psi_2$ is detrimental to one and beneficial to the other.
\end{itemize}
Therefore we are both level-SP and $L_1$-SP. 
\endproof

\printbibliography
%\bibliographystyle{informs2014} % outcomment this and next line in Case 1
%
%\bibliography{MORlevel-SP.bib} % if more than one, comma separated

\end{document}